\documentclass[useAMS,usenatbib]{mnras}

\DeclareSymbolFont{cmletters}{OML}{cmm}{m}{it}
\DeclareMathSymbol{v}{\mathalpha}{cmletters}{"76}

\usepackage{tabularx,ragged2e,booktabs,caption}
\usepackage{amsmath}
\usepackage{amssymb}
\usepackage{epsfig}
\usepackage{multicol}
\usepackage{graphicx}
\usepackage{ifthen}
\usepackage{latexsym}
\usepackage{rotating}
\usepackage{morefloats}
\usepackage{times,epsf}
\usepackage{txfonts}
\usepackage{varioref}
\usepackage{verbatim}
\usepackage{array}
\usepackage{url}
\usepackage{color}
\usepackage[dvipsnames]{xcolor}
\usepackage[T1]{fontenc}
\usepackage{epstopdf}
\usepackage{ulem}

\usepackage{tabularx,ragged2e,booktabs,caption}
\usepackage{amsmath}
\usepackage{amssymb}
\usepackage{epsfig}
\usepackage{ifthen}
\usepackage{latexsym}
\usepackage{rotating}
\usepackage{times,epsf}
\usepackage{txfonts}
\usepackage{varioref}
\usepackage{verbatim}
\usepackage{array}
\usepackage{url}
\usepackage{color}
\usepackage[dvipsnames]{xcolor}
\usepackage[T1]{fontenc}
\usepackage{epstopdf}
\usepackage{ulem}
\usepackage{graphicx}

\newcommand{\be}{\begin{equation}}
\newcommand{\ee}{\end{equation}}
\newcommand{\bea}{\begin{eqnarray}}
\newcommand{\eea}{\end{eqnarray}}

\title[Luminosity of ULX bubbles]{Optical and X-ray luminosity of expanding nebulae around ultraluminous X-ray sources}
\author[M. Siwek et al.]
       {Magdalena Siwek$^1$, Aleksander S\k{a}dowski$^{2,3}$, Ramesh Narayan$^4$, Timothy P. Roberts$^5$ \and and Roberto Soria$^{6,7}$ \thanks{E-mail:2028104m@student.gla.ac.uk (MM), asadowsk@mit.edu (AS)} \\        
$^1$ School of Physics \& Astronomy, University of Glasgow, G12 8QQ, UK \\
$^2$ MIT Kavli Institute for Astrophysics and Space Research,
77 Massachusetts Ave, Cambridge, MA 02139, USA\\
$^3$ Einstein Fellow\\
$^4$ Harvard-Smithsonian Center for Astrophysics,
60 Garden Street, Cambridge, MA 02134, USA\\
$^5$ Centre for Extragalactic Astronomy, Durham University, Department of Physics, South Road, Durham DH1 3LE, UK\\
$^6$ International Centre for Radio Astronomy Research, Curtin University, GPO Box U1987, Perth, WA 6845, Australia\\
$^7$ Sydney Institute for Astronomy, School of Physics A28, The University of Sydney, NSW 2006, Australia}

\begin{document}

\maketitle

\label{firstpage}

\begin{abstract}
We have performed a set of simulations of expanding, spherically
symmetric nebulae inflated by winds from accreting black holes in
ultraluminous X-ray sources (ULXs). We implemented a realistic
cooling function to account for free-free and bound-free cooling. For
all model parameters we considered, the forward shock in the
interstellar medium becomes radiative at a radius $\sim100$\,pc. The
emission is primarily in the optical and UV, and the radiative
luminosity is about 50\% of the total kinetic luminosity of the
wind. In contrast, the reverse shock in the wind is adiabatic so long
as the terminal outflow velocity of the wind $v_{\rm w}\gtrsim
0.003c$. The shocked wind in these models radiates in X-rays, but with
a luminosity of only $\sim10^{35}\,{\rm erg\,s^{-1}}$.  For wind
velocities $v_{\rm w}\lesssim 0.001c$, the shocked wind becomes
radiative, but it is no longer hot enough to produce X-rays. Instead
it emits in optical and UV, and the radiative luminosity is comparable
to 100\% of the wind kinetic luminosity. We suggest that measuring the
optical luminosities and putting limits on the X-ray and radio
emission from shock-ionized ULX bubbles may help in estimating the
mass outflow rate of the central accretion disk and the velocity of
the outflow.
 
\end{abstract}


\section{Introduction}
\label{s.introduction}

Ultraluminous X-ray sources (ULXs; see \citealt{Feng11} for a recent
review) are the most luminous persistent point X-ray sources located
outside the nuclei of galaxies.  Although some may host
intermediate-mass black holes with masses $\sim 10^2 - 10^5
~M_{\odot}$ \citep{farrell09,davis11,mezcua15}, the majority are now
thought to be less massive compact objects that accrete at extreme
rates (e.g. \citealt{gladstone09,sutton13}; \citealt{liu13,motch14}),
with at least 3 objects now confirmed to be neutron stars on the basis
of displaying X-ray pulsations \citep{bachetti14,israel16,fuerst16}.

Optical observations of ULXs reveal bubble-like nebulae (henceforth
ULX bubbles --- ULXBs), with diameters of up to 500 pc
(e.g. \citealt{pakull02,roberts03,ramsey06};
\citealt{abolmasov07,moon,cseh12}).  Based on standard diagnostic
optical line ratios ({\it e.g.}, [S\,II]/H$\alpha$, [N\,II]/H$\alpha$,
[O\,III]/H$\beta$), it was determined \citep{pakull02,pakull,moon}
that some of these nebulae contain mostly X-ray photo-ionized gas,
while others result mostly from shock-ionization. X-ray photoionized
nebulae such as those around the ULXs in Holmberg II
\citep{kaaret1,pakull02} and NGC\,5408 \citep{kaaret2} constrain the
isotropic photon luminosity and collimation angle of the compact
source. Shock-ionized ULXBs expand into the interstellar medium (ISM)
at a speed of a few 100 km s$^{-1}$ over characteristic timescales of
up to $10^6$ yr \citep{pakull}. In this paper, we focus specifically
on shock-ionized ULXBs: we model their physical structure and
luminosity at different evolutionary stages.

The mechanical power required to inflate a ULXB can be estimated from
optical observations in two independent ways \citep{Pakull2010}:
either from the flux emitted in specific diagnostic lines ({\it
  {e.g.}}, H$\beta$ and [Fe\,II] $\lambda$1.64$\mu$m), or from their
size and expansion speed (with plausible assumptions on the ISM
density). Both methods suggest that ULXBs contain at least one order
of magnitude more energy than an ordinary supernova remnant, as well
as being much larger and longer-lived. Several formation scenarios
have been proposed.

In one scenario, the bubbles are inflated by outflows from multiple O
stars and supernova remnants, forming a so-called "superbubble" An
alternative scenario is that they result from "hypernova" events,
which may inject a large amount of energy (up to $\sim$10$^{53}$ {\rm erg})
in a single explosion. However, both of these formation scenarios are
disfavored by observations \citep{pakull, pakull2005}. In most cases,
the stellar population inside and around ULXBs is moderately old
($\approx$20 Myr), does not contain stars massive enough to produce
hypernovae at the current epoch, and does not contain enough OB stars
to produce superbubbles.

A more likely scenario is that ULXBs are formed by the continuous
injection of winds/jets from an accretion disk surrounding a black
hole or neutron star at the centre of the ULX.  Interestingly, this is
qualitatively consistent with the super-Eddington models that have
been invoked to explain many of the X-ray characteristics of ULXs
\citep{poutanen07,rs3,middleton15,narayan17}, where the extreme
radiation release from the central regions of the super-Eddington
accretion flow drives a massive wind away from the accretion disc;
such a wind may be a prime culprit for the inflation of the ULXBs.  In
this model, the wind would need to input $\sim 10^{39}{\rm erg\,s^{-1}}$
into the nebula to inflate it, meaning the mechanical output of ULX
disks must be similar to their radiative output
\citep{pakull02,roberts03}.  Evidence for this wind has recently
emerged from X-ray observations, with the detection of absorption
features from a medium that is outflowing at a velocity $\sim 0.2c$ in
the high resolution X-ray spectra of two ULXs \citep{pinto}.
Rest-frame emission lines are also detected that might be attributable
to collisional heating in the vicinity of the ULX, although, with the
current quality of data, photoionization models also provide a
plausible explanation for these features. Meanwhile, general
relativistic radiation MHD simulations of super-Eddington black hole
accretion have also confirmed that such systems invariably produce
powerful winds \citep{sadowski+koral2,rs2,narayan17}.

In this paper we investigate the ULX disk wind model in detail. We
perform numerical simulations of expanding nebulae with realistic
cooling and track the evolution of the shocked gas. We investigate the
properties of the shocked wind and the related optical and X-ray
emission as a function of the disk outflow parameters. 
%We find that, while the integrated wind luminosity in X-ray grows
%with decreasing outflow velocity, below some threshold of the
%velocity the layer of the shocked wind becomes radiatively efficient
%and cools primarily in optical/UV, similarly to the shocked ISM.
We suggest that optical and X-ray (and potentially radio) emission
properties of the ULX nebulae may constrain the characteristics of the
outflows blown out of the accreting systems.

The paper is structured as follows. We first describe the numerical
methods and the setup of our simulations (Section~\ref{s.methods}). We
then discuss the properties of the expanding nebula in the fiducial
model (Section~\ref{s.fiducial}), followed by a parameter study where
we investigate the effect of wind velocity on the nebular properties
(Section~\ref{s.parameter}). Caveats and astrophysical implications
are discussed in Section~\ref{s.discussion}, and the results are
summarized in Section~\ref{s.summary}.

\section{Numerical methods}
\label{s.methods}

\subsection{\texttt{KORAL}}
\label{s.koral}

In the scenario investigated in this paper, the expanding nebulae in
ULXBs result from kinetic power pumped out by a central accreting
black hole (BH) via a quasi-spherical wind. The outflowing gas pushes
and shocks the ISM. The nebula is optically thin, but may cool
significantly enough to affect its dynamics.

The simulations described here were carried out with the code
\texttt{KORAL} \citep{sadowski+koral, sadowski+koral2}, which is
capable of evolving magnetized gas and radiation in parallel in a
relativistic framework, for arbitrary optical depths. In the project
described here, we neglect magnetic fields, the gravity of the BH
(since the interesting interaction of the outflow with the ISM takes
place at large radii), and radiative transfer (since the gas is
optically thin throughout the expansion). We also assume isotropic
expansion of the nebula and evolve the problem in one (radial)
dimension.

\subsection{Cooling function}
\label{s.cooling}

\begin{figure}
\centering
\includegraphics[width=1\linewidth]{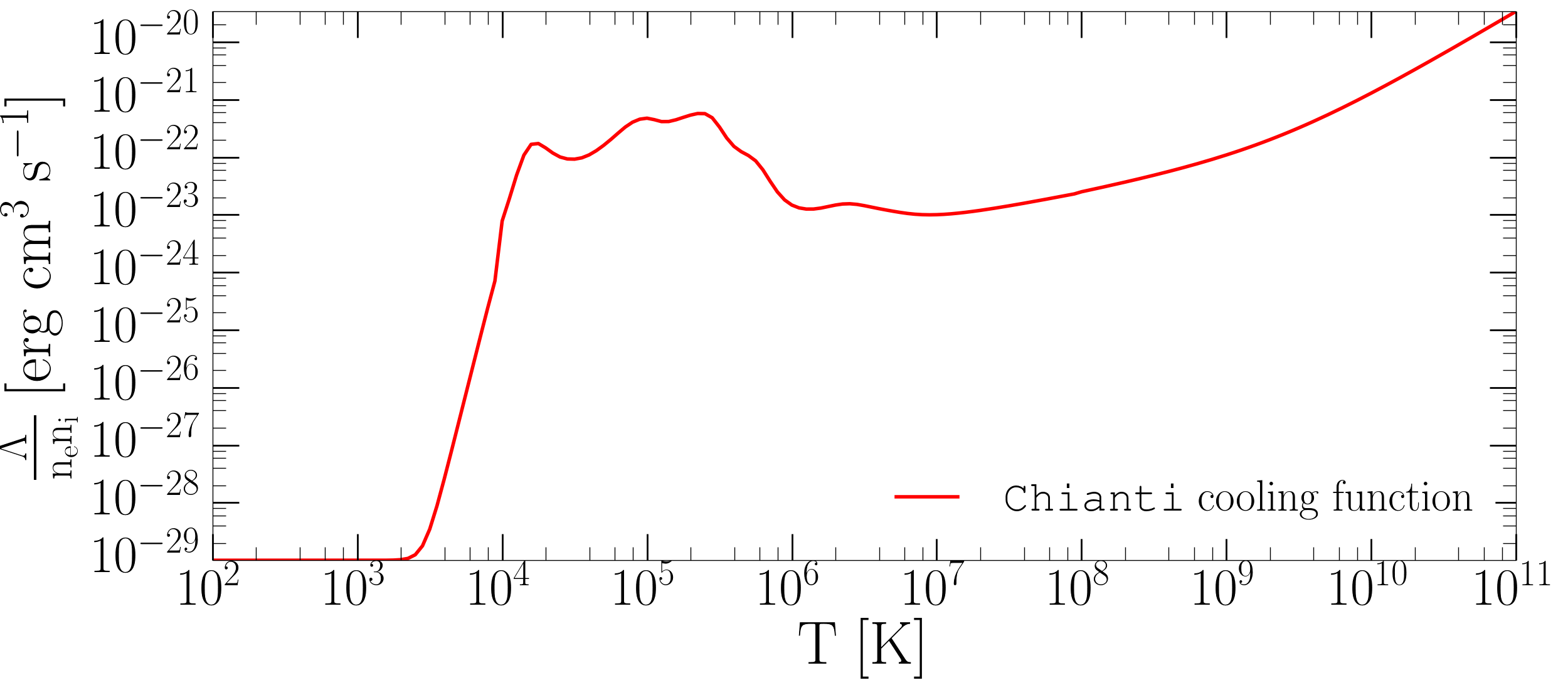}
\caption{Gas emissivity $\Lambda$ for solar abundances \color{black} as a function
  of temperature $T$, obtained from the CHIANTI Database Version 8.
  Above $T = 10^{8}K$, the emissivity is modelled as a broken
  power-law to account for relativistic bremsstrahlung (equation
  \ref{brems}).}
\label{fig:temp_vs_cooling_2}
\end{figure}

To account for bound-free and free-free cooling we provide
\texttt{KORAL} with opacities calculated from a cooling function
generated with \texttt{ChiantiPy}, the \texttt{Python} interface to
the \texttt{Chianti} database \citep{chianti1,chianti8}. Figure
\ref{fig:temp_vs_cooling_2} shows the radiative loss rate we use in
the present study (obtained with the \texttt{Chianti RadLoss}
function, assuming solar abundances).

Although the cooling rate falls off naturally at temperatures of order
$10^4$\,K or below, we reduce it further to ensure that the
unperturbed ISM, which is at an equilibrium temperature of $10^4$\,K
through cosmic ray heating, does not cool significantly during the
course of the simulation. We leave the cooling function obtained
through \texttt{ChiantiPy} unmodified between $10^4$ K and $10^8$ K,
and then extrapolate up to $10^{11}$K assuming pure free-free
emission. Including a relativistic correction
\citep{sadowski+electrons}, we assume the following broken power-law
for the cooling function at high temperatures,
\begin{equation}
\label{brems}
 \frac{\Lambda_{\rm ff} \color{black}}{n_{\rm i} n_{\rm e}} = 2.409\times 10^{-27}
\sqrt{T}\left(1+4.4\times 10^{-10}T\right) \ {\rm erg\,cm^3\,s^{-1}}, \quad
T>10^8\,{\rm K}.
\end{equation}

The above opacities are used when calculating the source terms that
describe energy loss of the gas due to radiation
\citep{sadowski+koral}. Since the gas is optically thin, no radiative
transfer calculation is done as part of the simulation.

\subsection{Numerical setup}
\label{s.setup}

Current observations \citep{pakull, pakull2005} suggest that ULXBs are
formed by the continuous injection of winds/jets from the accretion
disk surrounding the black hole at the centre of the ULX.  The
underlying physics has been extensively studied, most notably by
\cite{castor} and \cite{weaver}, in the context of interstellar
bubbles formed by the interaction of stellar winds with the ISM.

The simulations are set up as follows. We introduce an isotropic wind
which has been emitted from the accretion disk but has since reached
some constant velocity $v_w$ far from the disk. This wind expands and
propagates freely into the ISM until its density is equal to that of
the ambient medium.  The radius at which this happens is called the
initial contact surface (CS) and depends on the kinetic wind
luminosity $ L_w$, wind velocity $ v_{\rm w}$ and the density of the
surrounding ISM $\rho_{\rm ISM}$.

For a fixed kinetic luminosity and wind velocity, the density of the
wind $\rho_w$ varies with radius $r$ as
\begin{equation}
\label{windprofile}
{\rho_{\rm w} = \frac{L_{\rm w}}{2 \pi r^2 v_{\rm w}^3}}.
\end{equation}
Equating this to $\rho_{\rm ISM}$, the location of the CS is given by
\begin{equation}
\label{cs}
{r_{\rm CS} = \sqrt{\frac{L_{\rm w}}{2 \pi \rho_{ \rm ISM} v_{\rm w}^3}}}.
\end{equation}
We begin each simulation with the wind filling the sphere up to radius
$ r_{\rm CS}$, with an unperturbed constant density ISM beyond this
radius.  The start time of the simulation thus corresponds to the end
of the free expansion phase of the ULXB. Correspondingly, the physical
time since the outflow turned on is
\begin{equation}
 t_{\rm free} = \frac{r_{\rm CS}}{v_{\rm w}}.
\end{equation}

The temperature of the injected wind is held constant at $ T_{\rm
  wind} = 10^6$ K.  Typical ISM densities surrounding ULXs are of
order $\rm 1\, \rm{cm^{-3}}$ \citep{pakull}, the typical density range of
the Warm Ionized Medium at temperature $ T \approx 10^4 $K
\citep{draine}. We adopt these values for the ISM in all the
simulations.

\begin{table}
\begin{center}
\caption{Model parameters}
\label{tab:ode}
\begin{tabular}{lllll}
\hline
\hline
Name &
$M_{\rm BH}$ & $L_{\rm w}$ & $v_{\rm w} $ & $r_{\rm CS}$ \\
\hline
\hline
\texttt{L40v-1.5} & $10M_\odot$ & $10^{40}\,\rm erg s^{-1}$ & $10^{-1.5}c$ & $0.34\,\rm pc$\\
\textbf{\texttt{L40v-2}} & \textbf{$10M_\odot$} & \textbf{$10^{40}\,\rm erg s^{-1}$} & \textbf{$10^{-2}c$} &  \textbf{$1.9\,\rm pc$}\\
\texttt{L40v-2.5} & $10M_\odot$ & $10^{40}\,\rm erg s^{-1}$ & $10^{-2.5}c$ & $10.8\,\rm pc$\\
\texttt{L40v-3} & $10M_\odot$ & $10^{40}\,\rm erg s^{-1}$ & $10^{-3}c$  & $60.9\,\rm pc$\\
\texttt{L39v-2} & $10M_\odot$ & $10^{39}\,\rm erg s^{-1}$ & $10^{-2}c$ & $0.61\,\rm pc$\\
\hline
\hline
\multicolumn{5}{l}{$M_{\rm BH}$ -- mass of the BH, $L_{w}$ - mechanical luminosity}\\
\multicolumn{5}{l}{of the outflow, $ v_{\rm w} $ -- velocity of the outflow,}\\
\multicolumn{5}{l}{$ r_{\rm CS}$ - radius at which ISM density starts to dominate.}\\
\multicolumn{5}{l}{The fiducial model (\texttt{L40v-2}) is highlighted.}\\

\end{tabular}
\end{center}
\end{table}

We keep the mechanical luminosity of the central BH engine constant at
either ${L_{\rm w} = 10^{40} \, {\rm erg\,s^{-1}}} $ (in most cases) or ${L_{\rm
    w} = 10^{39}{\rm erg\,s^{-1}}}$ (for one simulation), and vary the wind
velocity from ${v_{\rm w} = 10^{-3}c}$ to ${v_{\rm w}= 10^{-1.5}c}$
(see Table~\ref{tab:ode})\footnote{Assuming that the outflow reaches
  infinity with a velocity equal to a reasonable fraction of the
  Keplerian velocity at the launch radius in the accretion disk, these
  velocities correspond to a wide range of launch radii. We do not
  consider the highest-velocity outflow from the inner region of the
  disk, since that is likely to emerge as a jet rather than the
  quasi-spherical wind considered in this paper. The interaction of a
  jet with the ISM requires at least 2D (axisymmetric) simulations,
  which we leave for future work.}.  We investigate how the wind
parameters affect the dynamical evolution of the shock and the
radiative properties of the ULXB. In particular, we examine the
radiative emission of the ULXB during the adiabatic and radiative
phases in X-rays, optical and radio, and its dependence on the wind
velocity.

Most of the simulations are run at a resolution $N_{\rm R} = 4608$,
i.e. 4608 cells over a range of radii between $ r_{\rm CS}$ and
500\,pc, spaced logarithmically. However, we run the fiducial model ($
L_{\rm w} = 10^{40}{\rm erg\,s^{-1}}$, $ v_{\rm w} = 0.01c$, model
\texttt{L40v-2}) additionally at two higher resolutions: $ N_{\rm R} =
9216$ and $ N_{\rm R} = 13824$. This allows us to study the effect of
resolution on the radiative properties at the contact discontinuity
(CD) and the forward shock.

\section{Simulations of ULX Nebulae}
\label{s.results}

\subsection{Fiducial model -- \texttt{L40v-2}}
\label{s.fiducial}

\subsubsection{Expansion}

In the classical picture of wind inflated bubbles, the emitted wind
expands freely for a short while until its density is comparable to
the density of the ambient medium. This is the free expansion
phase. \cite{weaver} describe the next stages of bubble expansion as
follows. As the wind reaches the critical radius $ r_{\rm CS}$
(equation \ref{cs}), a reverse and forward shock start to
develop. However, the shocks expand fast enough so that the shocked wind
and shocked ISM do not cool and the flow remains adiabatic. This is
the adiabatic phase, in which the forward shock propagates outward
with radius increasing with time as
\begin{equation}
\label{fs} R_2 = \alpha_1 \left(\frac{L_{\rm w}t^3}{\rho_{\rm ISM}}\right)^{0.2},
\end{equation}
where $ \alpha_1$ is a dimensionless constant $\approx 0.88$
\citep{weaver}.

\begin{figure}
\centering
\includegraphics[width=1\linewidth]{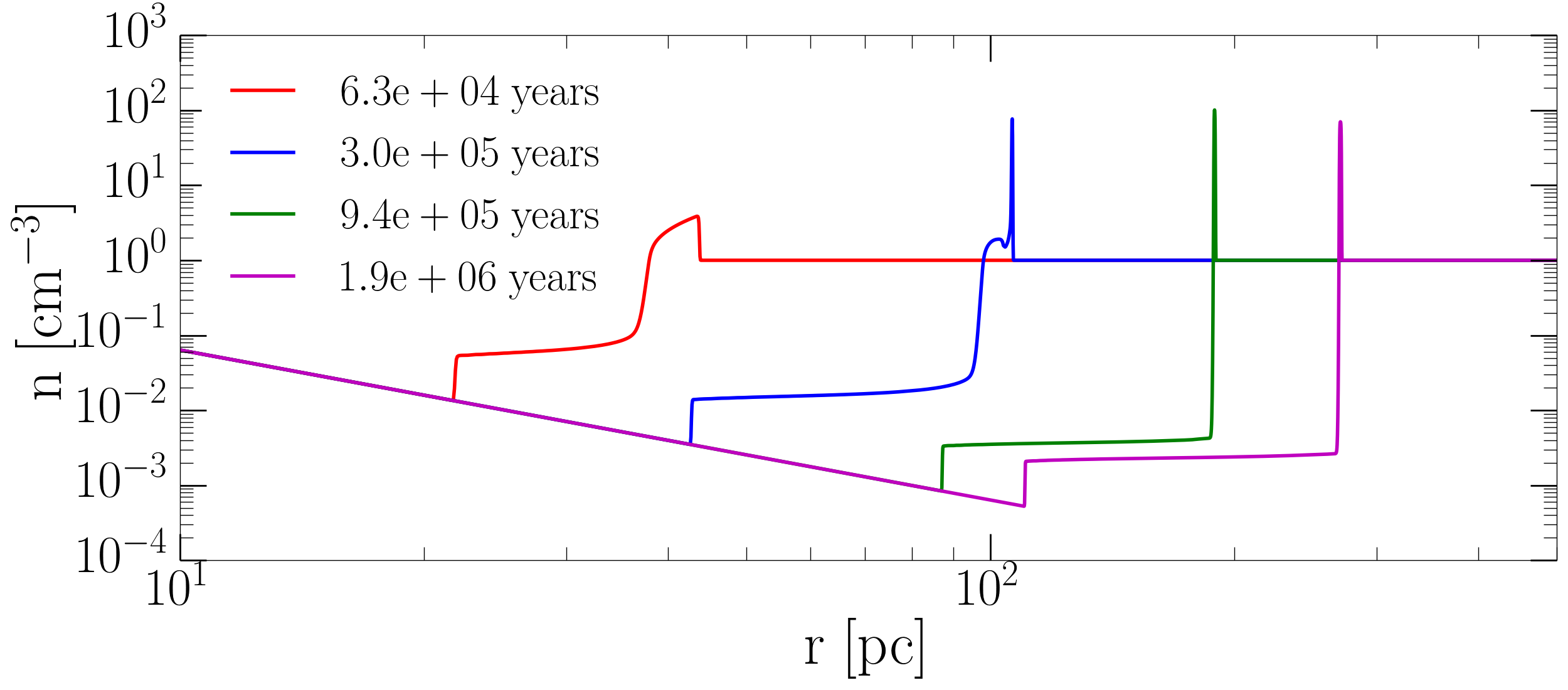}
\includegraphics[width=1\linewidth]{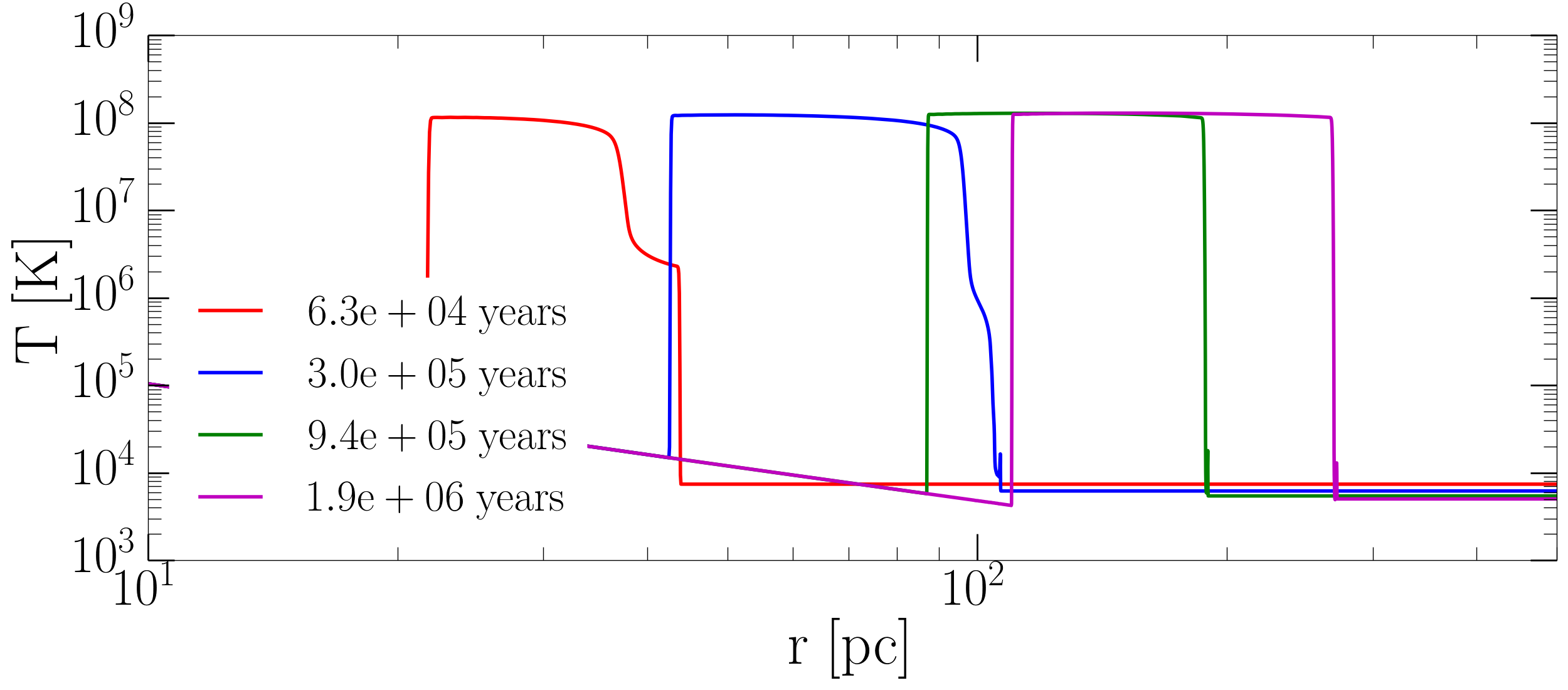}
\includegraphics[width=1\linewidth]{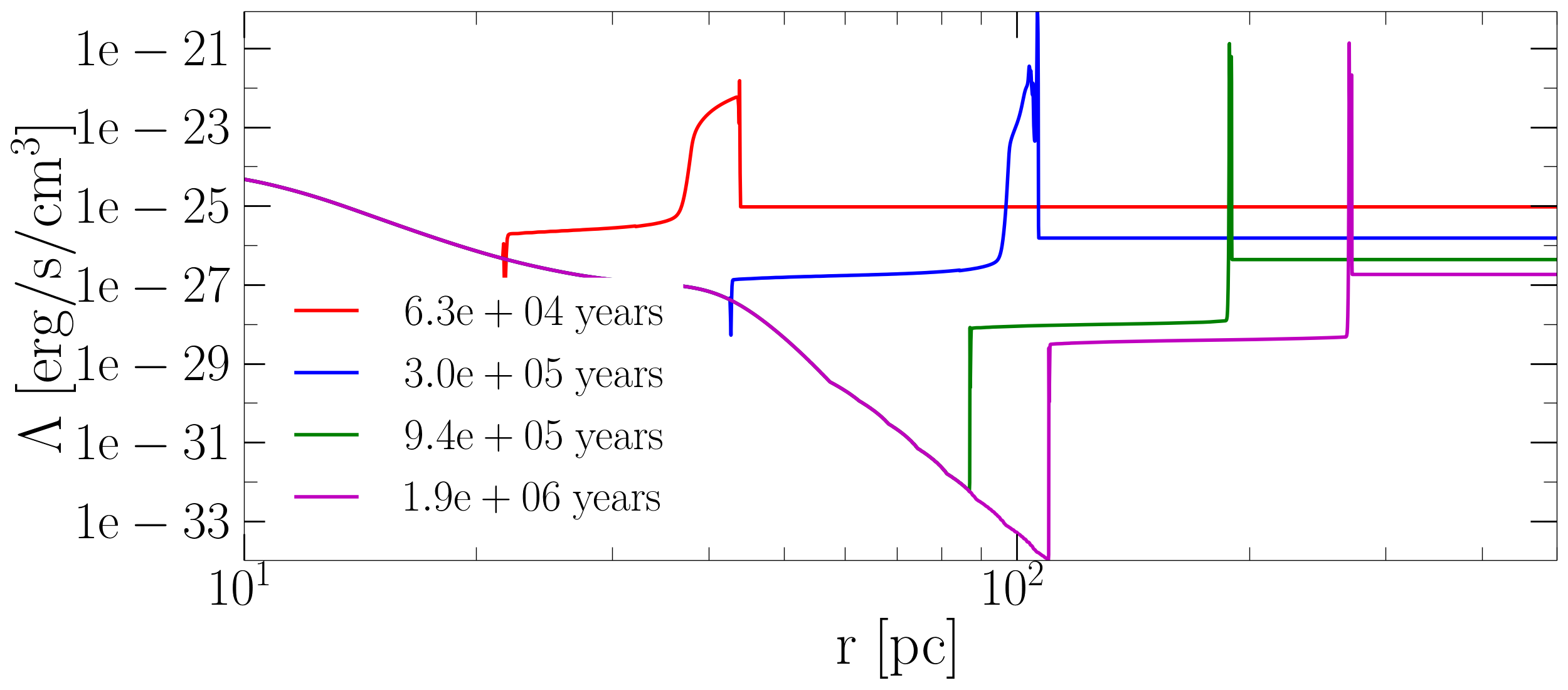}
\includegraphics[width=1.005\linewidth]{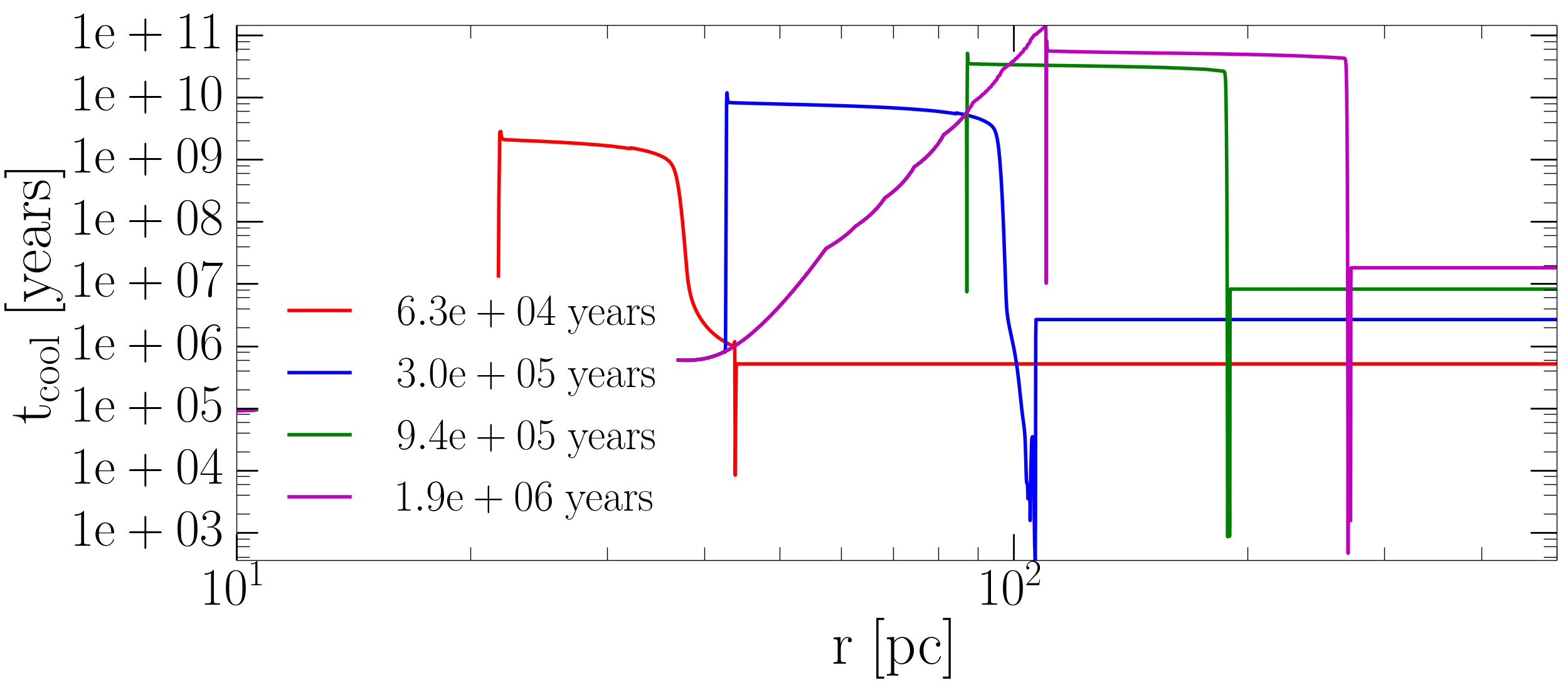}
\caption{Panels 1-4 respectively show profiles of density,
  temperature, cooling rate and cooling time scale for the expanding
  bubble in the fiducial model \texttt{L40v-2}. Four different times
  are shown. The red line at $t = 6.3 \times 10^4 $ years shows an
  extended layer of shocked ISM, which begins collapsing into a thin
  shell at $t = 3 \times 10^5 $ years (blue line). This marks the
  transition from the adiabatic to the radiative stage of expansion,
  where the shocked ISM begins to cool efficiently. The shocked wind
  remains adiabatic and continues to expand as a thick envelope of
  tenuous hot gas until late times.}
\label{fig:L40v-2}
\end{figure}

Since the shocked ISM is adiabatic, we expect to see an extended
envelope of hot, somewhat tenuous gas at the front shock. Indeed, the
early phase of the expansion in the fiducial model follows this
picture: the red lines in Fig.~\ref{fig:L40v-2} show an extended
envelope of shocked ISM with $ n \approx 4\, \rm{cm^{-3}}$ (top panel) and
a temperature of several $10^{6}$\,K (second panel from the top). The
bottom two panels show the cooling rate $\Lambda$ and the cooling time
scale $t_{\rm cool}$, which is defined as the ratio of the gas
internal energy $\rm u_{\rm int}$ and the cooling rate,
\begin{equation}
 t_{\rm cool} = \frac{u_{\rm int}}{\Lambda} .
\end{equation}
At the location of the forward shock, $ t_{\rm cool}$ exceeds the age
of the ULXB during the adiabatic phase.

When the age of the bubble becomes comparable with the cooling time of
the shocked ISM, i.e., when the shocked ISM starts to be radiatively
efficient, radiative losses occur in the forward shock and the swept
up ISM collapses into a thin shell. The transition to this phase is
seen in the blue lines of Figure \ref{fig:L40v-2}. The front edge of
the shocked ISM has already begun to collapse to a thin, dense
shell. The temperature profile shows the corresponding cooling of the
very same gas. This is consistent with the cooling time scale in the
bottom panel, which, at the location of the front shock, is now far
less than the age of the ULXB.  In the radiative phase the shocked ISM
cools down to temperatures below $10^4$\,K at which further cooling is
suppressed (see the cooling function in Figure
\ref{fig:temp_vs_cooling_2}). However, newly shocked ISM which is
swept up as the ULXB expands is heated rapidly behind the front shock
and reaches temperatures of several $10^5$K (second panel of Figure
\ref{fig:L40v-2}), and only then cools rapidly. This can be
seen in the troughs in the bottom panel, indicating the short
cooling time due to efficient bound-free emission. These properties
hold for the rest of the expansion (compare green and magenta lines in
Fig.\ref{fig:L40v-2}) which proceeds with slightly slower propagation
speed.  The expansion rate when the shocked ISM is radiative still
follows the same dimensional dependences as in equation \ref{fs}.
However, according to \cite{weaver}, the factor $\alpha_1$ is replaced
by $\alpha_2 \approx 0.76$:
 \begin{equation}
 \label{fs2}
 {R_2 = \alpha_2 \left(\frac{L_{\rm w}t^3}{\rho_{\rm ISM}}\right)^{0.2}}.
 \end{equation}

The above discussion pertains to the forward shock. The reverse shock
is adiabatic throughout the expansion, as long as the outflow is fast
enough so that the cooling time of the shocked wind is long. This is
true for all the simulated models except \texttt{L40v-3}, where the
cooling time in the shocked wind is comparable to that of the shocked
ISM. Therefore, in model \texttt{L40v-3}, both layers become radiative
at the same time.

\subsubsection{Radiative properties}

A bolometric light curve can be calculated directly from
\texttt{KORAL} by summing at each instant of time the total cooling
over all spherical shells in the ULXB.  To calculate light curves in
specific frequency bands we need to calculate spectra as a function of
time, which we do as follows.

We take the density and temperature profiles directly from the
\texttt{KORAL} simulation, and estimate the cooling rate $\Lambda$ for
the gas in each cell using Figure \ref{fig:temp_vs_cooling_2}. For the
spectrum emitted from a shell of width $ \delta R$ at radius $ R$
and temperature $ T$, we assume for simplicity that it has the same
shape as bremsstrahlung emission and write
\begin{equation}
S_{\rm \nu} = {\frac{h\Lambda}{kT} \exp\left({-
    \frac{h\nu}{kT}}\right)}\times 4\pi R^2 \delta R.
\label{eq:Snu}
\end{equation}
We note that by normalizing the spectrum with $\Lambda$, the cooling rate obtained by \texttt{CHIANTI}, we include emission via bound-free cooling as well as free-free cooling. Therefore, even though the shape is approximated to a bremsstrahlung spectrum, the overall emission rate is preserved. This is significant because the difference between the cooling rates spans several orders of magnitude for gas at temperatures $\lessapprox 10^6$\,K, where most of the radiation is emitted as metal lines. By preserving the total emission rate, our bremsstrahlung approximation ought to provide an order-of-magnitude approximation of the integrated luminosities in various bands (X-ray, optical, radio), but it is not suitable for detailed spectral comparisons. In future work we could improve on our spectrum approximation by calculating the line spectrum of the shock-ionized gas in each cell using the \texttt{APEC} code \citep{rs7}.

Using our bremsstrahlung approximation, we integrate the spectra over all shells in the shocked ISM and wind, and estimate the total spectrum of the ULX bubble. This direct method of calculating spectra is adequate during the early adiabatic stage of the bubble. Luminosities in individual bands are calculated by integrating the ULXB spectrum over the corresponding frequency ranges. \color{black}

During the radiative phase, the emission at the front shock is
dominated by rapid cooling of the newly swept up and shocked ISM.
However, as we discuss in \S\ref{s.resolution}, numerical simulations
have trouble resolving the temperature structure of radiative
shocks. Typically, when the radiative efficiency of the shocked gas is
high, the post-shock gas in the simulation does not achieve the
correct temperature but ends up somewhat cooler because of the finite
cell size. If we used the simulation data to calculate the spectrum,
we would obtain a misleadingly soft spectrum, with its peak at a lower
frequency. This is an issue during the radiative phase of the
expansion, when the emission is dominated by the front
shock. Fortunately, radiative shocks can be treated analytically to
compute the correct spectrum (the following discussion follows
\citealt{draine}).

Consider a spherical shock at radius $R$ which moves outward with a
radial velocity $v_{\rm sh}$ into a homogeneous ISM with density
$\rho_{\rm ISM}$. The rate
at which the mass of shocked ISM increases with time is given by
\begin{equation}
\dot{M}_{\rm sh} = 4\pi R^2 v_{\rm sh}\,\rho_{\rm ISM}.
\end{equation}

Each parcel of gas is hottest immediately after the shock, where its
temperature is given by the adiabatic shock jump conditions. For
simplicity, we assume that the external medium is ``cold'', which means
that the shock velocity is much greater than the sound speed in the
medium. This is true for ULX bubbles, since the sound speed in the
medium is only about $10~{\rm km\,s^{-1}}$, whereas the shock moves at
hundreds of $\rm km\,s^{-1}$. Therefore, the immediate post-shock
density and pressure are given by,
\begin{equation}
\rho_2 = \frac{\gamma+1}{\gamma-1}\,\rho_{\rm ISM} = 4\rho_{\rm ISM},
\end{equation}
\begin{equation}
  P_2 = \frac{2}{\gamma+1}\, \rho_{\rm ISM} v_{\rm sh}^2 = \frac{3}{4}\,\rho_{\rm ISM}v_{\rm sh}^2,
\end{equation}
where we have set the adiabatic index $\gamma = 5/3$. The immediate post-shock temperature is then
\begin{equation}
\label{t2}
T_2 = \frac{\mu m_{\rm p} P_2}{k\rho_2} = \frac{3}{16}\, \frac{\mu m_{\rm p}}{k}\,
v_{\rm sh}^2,
\end{equation}
where $\mu$ is the effective molecular weight of the gas.

Each parcel of gas starts with temperature $\rm T2$ (and density
$\rho_2$) immediately after it crosses the shock, but it then cools
down to a final temperature $\rm T_3$ which it reaches far
downstream. In our simulations, $T_3$ is typically between $4000 -
8000$\,K. The cooling from $T_2$ to $T_3$ converts thermal energy to
radiation, and we can calculate how much radiation is emitted at each
temperature.

Immediately post-shock, the downstream gas density is $\rho_2=4\rho_{\rm ISM}$
and its velocity is $u_2=v_{\rm sh}/4$.  However, both $\rho(r)$ and
$u(r)$ vary with distance away from the shock because cooling causes
the temperature $T(r)$ to vary. Since the mass flux is conserved in
the shock frame,
\begin{equation}
\rho(r) u(r) = \rho_{\rm ISM} v_{\rm sh} ~~\to~~ \frac{\rho(r)}{\rho_{\rm ISM}} =
\frac{v_{\rm sh}}{u(r)} \equiv x(r),
\end{equation}
where $ x(r)$ measures the density compression at any given radius
r downstream of the shock.  (We ignore the spherical geometry in this
analysis since the cooling region of the radiative shock is quite thin
compared to the local radius.) Momentum flux is also
conserved. Ignoring the upstream pressure, this gives
\begin{equation}
\rho(r) u^2(r) + P(r) = \rho_{\rm ISM} v_{\rm sh}^2 ~~\to~~ P(r) =
\frac{[x(r)-1]}{x(r)}\, \rho_{\rm ISM} v_{\rm sh}^2.
\end{equation}
Rewriting in terms of the temperature $T(r) = \mu m_{\rm p} P(r) / k
x(r) \rho_{\rm ISM}$, we obtain the following quadratic equation for
$x(r)$ as a function of $T(r)$:
\begin{equation}
\frac{kT(r)}{\mu m_{\rm p}}\,x^2(r) - v_{\rm sh}^2\, x(r) + v_{\rm sh}^2 = 0.
\label{quad}
\end{equation}
The solution to this equation is
\begin{equation}
x(r) = \frac{\mu m_{\rm p}}{2kT(r)}\left[v_{\rm sh}^2 + \left(v_{\rm sh}^4 - 
4 v_{\rm sh}^2 \frac{kT(r)}{\mu m_{\rm p}}\right)^{1/2}\right],
\end{equation}
with
\begin{equation}
\frac{dx}{dT} = \frac{x^2(r)}{v_{\rm sh}^2 - 2x(r)kT(r)/\mu m_{\rm p}}.
\end{equation}

Consider now the energy equation of the shocked gas and the radiative
luminosity. The differential amount of radiation $dE_{\rm rad}$
emitted by unit mass of the gas as its temperature changes from
temperature $T$ to $T+dT$ is
\begin{equation}
-dE_{\rm rad} = dQ = du + P(r)d\left[\frac{1}{\rho(r)}\right]
= \frac{k}{\mu m_{\rm p}}\left[\frac{3}{2} - \frac{T(r)}{x(r)} \frac{dx}{dT}
\right] dT.
\end{equation}
Substituting for $x$ and $dx/dT$, and scaling up to the
entire volume of the shocked gas, we obtain the total luminosity
emitted in a logarithmic temperature interval $\Delta \ln T$,
\begin{equation}
\Delta \dot{U} = 4\pi R^2 v_{\rm sh}\rho_{ISM} \,
\frac{kT}{\mu m_{\rm p}}\,\left[\frac{5}{2} + 
\frac{v_{\rm sh}^2 - (v_{\rm sh}^4 - 4v_{\rm sh}^2 kT/\mu m_{\rm p})^{1/2}}
{(v_{\rm sh}^4 - 4v_{\rm sh}^2 kT/\mu m_{\rm p})^{1/2}} \right]\, \Delta \ln{T}.
\label{dudlnt2}
\end{equation}

As before, to calculate the spectrum of this radiation, we assume that
the heated gas emits with a spectrum similar to bremsstrahlung
radiation.  Thus, following equation (\ref{eq:Snu}), we write the
spectral emission from gas at temperature $T$ as
\begin{equation}
\label{Lnu}
\frac{ dL_{\nu}^{ff}}{d\ln T} = 
\frac{\Delta \dot{U}}{\Delta \ln T} \, \frac{h}{kT}
\exp\left(-{\frac{h \nu }{k T}}\right).
\end{equation}
Summing up all the spectra over the temperature range from $\rm T_3$
to $\rm T_2$, we obtain the net spectrum of the forward shock region
in the ULXB.

To calculate the spectrum during the radiative phase,
we first use directly the density and temperature profiles
from the simulation to obtain the spectrum from the shocked wind and
the cooled region of the shocked ISM. We then define a cutoff at the
temperature minimum of the swept up ISM,
%(mid panel in Figure \ref{fig:all_res}) 
and for the region between this cutoff and the location of the forward
shock, we follow the analytical approach described above. The gas temperature at the
cutoff location is the final temperature $ T_3$ of the shocked ISM. We
calculate the immediate post-shock temperature $ T_2$ analytically
(equation \ref{t2}) by estimating the shock velocity from
\texttt{KORAL}. Then we integrate equation (\ref{Lnu}) from $T_2$ to
$T_3$ to calculate the analytical spectrum from the radiative forward
shock of the bubble.

\begin{figure}
\centering
\includegraphics[width=1.005\linewidth]{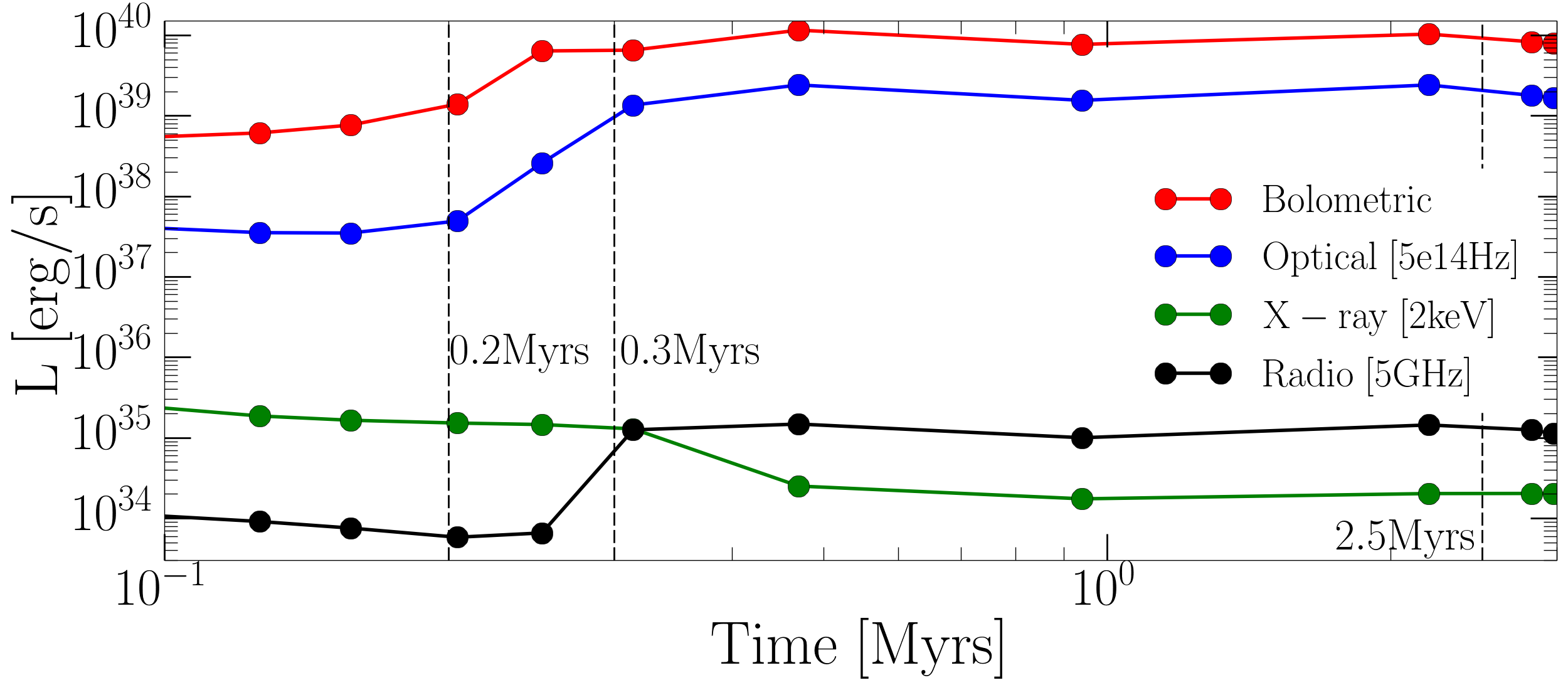} %\vspace{-.45cm}
\includegraphics[width=1\linewidth]{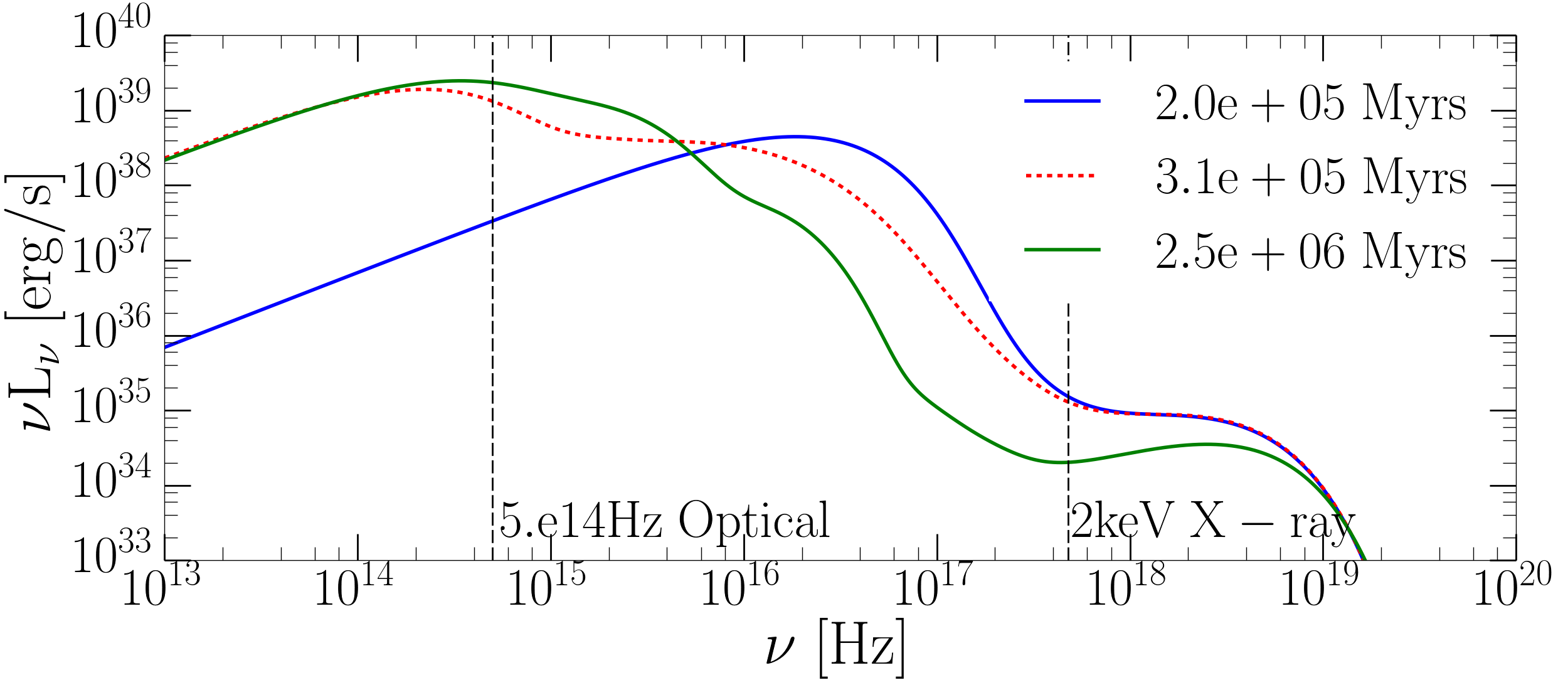}
\caption{Top: Volume-integrated emission of the ULX bubble in the
  fiducial run (model L40v-2, $L_{\rm w} = 10^{40} \, {\rm erg\,s^{-1}}$,
  $v_{\rm w} = 10^{-2}c$) as a function of time. Red, blue, black and
  green lines show bolometric, optical ($5\times10^{14}$\,Hz), radio
  (5\,GHz) and X-ray (2\,keV) light curves. Bottom: Spectra
  corresponding to the early adiabatic phase of the forward shock
  (blue), the transition from adiabatic to radiative (red dashed), and
  fully radiative (green line). }
\label{fig:1_e-2_4608_spectra-dashed_vertlines}
\end{figure}

In Figure \ref{fig:1_e-2_4608_spectra-dashed_vertlines} we show light
curves and spectra corresponding to the fiducial model
\texttt{L40v-2}.  The initial bolometric and optical luminosities are
low. After about $\approx 0.25 $\,Myrs, the bolometric luminosity
begins to increase, reaching $L_{\rm bol}\approx 10^{40} \, {\rm erg\,s^{-1}}$,
at 0.5\,Myrs. During the adiabatic phase at the start of expansion,
the wind luminosity is stored in the shocked wind and ISM. This energy
is emitted once the ULXB enters the radiative stage. When the
transition to the partially radiative phase is complete (at t $\approx
0.5\,$Myrs) the luminosities in all bands reach a stable value, roughly
50\% of the wind mechanical luminosity (comparable to the classical
value of 27/77 from bubble theory: Weaver et al.~1977). The radio
emission at 5\,GHz (black line) stays at $\approx 10^{34} \, {\rm erg\,s^{-1}}$ at
early times, and increases by an order of magnitude after 0.3\,Myrs.

The bottom panel of Fig.~\ref{fig:1_e-2_4608_spectra-dashed_vertlines}
shows computed spectra corresponding to the adiabatic, transition and
radiative phases. The adiabatic and transition spectra were obtained
purely from the density and temperature profiles from KORAL, whereas
the spectrum in the radiative phase is partially from KORAL but using
the previously described analytical model for the radiation from the
radiative forward shock.  The evolution of the spectrum shows that
X-ray emission from the ULXB is highest during the initial adiabatic
expansion. The bolometric luminosity, on the other hand, is low at
early times (as can be seen in the top panel of Figure
\ref{fig:1_e-2_4608_spectra-dashed_vertlines}), but the emission peak
is shifted towards higher frequencies.  By the time the transition to
the radiative phase is completed at $\approx 0.5$\,Myrs, the X-ray
emission has already fallen well below $10^{35} \, {\rm erg\,s^{-1}}$, which is
seen also in the spectrum at this time (green lines in the top and
bottom panels of Figure
\ref{fig:1_e-2_4608_spectra-dashed_vertlines}). At late times the peak
of the spectrum has shifted towards the optical/UV regime. The light
curves reflect this: While the X-ray luminosity has decreased
significantly, the optical emission remains stable at several
$10^{39}{\rm \,  erg\,s^{-1}}$ throughout the radiative phase.

\begin{figure}
\centering
\includegraphics[width=1\linewidth]{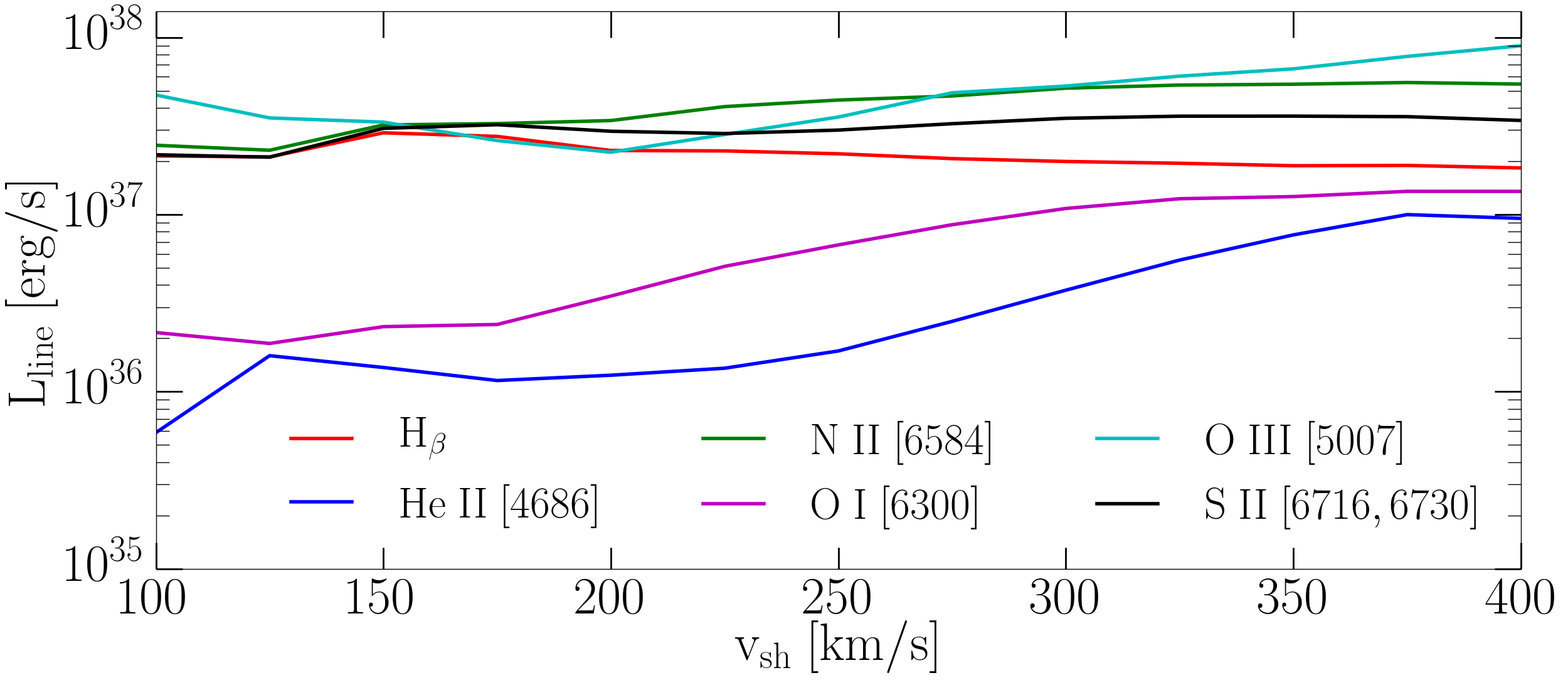}
\caption{Main diagnostic optical emission line luminosities
  (He\,II\,$\lambda 4686$, H$_{\beta}$, [O\,III]\,$\lambda 5007$,
  [O\,I]\,$\lambda 6300$, [N\,II]\,$\lambda 6584$ and the doublet
  [S\,II]\,$\lambda 6716,6730$) as a function of shock velocity,
  predicted for an input mechanical power of $10^{40}{\rm erg\,s^{-1}}$. The luminosities have been calculated with the
  shock-ionization code Mappings III (Allen et al.~2008), for an ISM
  density of 1 cm$^{-3}$, solar metallicity, and an equipartition
  magnetic field.}
\label{fig:lines}
\end{figure}

A pure bremsstrahlung approximation for the shape of the emitted spectrum provides a convenient
order-of-magnitude estimate for the shape of the emitted spectrum, especially for the X-ray and radio
bands, but it is inadequate to describe the optical spectrum from the
shocked ISM layer, for any realistic composition of the ISM. To
compare our results with optical observations, we need to insert our
model parameters (input mechanical power, $v_{\rm sh}$, ISM density)
into a shock-ionization code such as Mappings III
\citep{map1,map2,map3}. Table \ref{tab:shockvel} lists the velocity of
the forward shock at characteristic epochs for the various simulated
models. As an example, we have used these velocities to calculate and
plot (Figure \ref{fig:lines}) the predicted spectrum of important
diagnostic lines, for an input mechanical power of $10^{40} \, {\rm erg\,s^{-1}}$, ISM density of 1$ \, \rm{cm^{-3}}$, solar metallicity, and an
equipartition magnetic field. For some lines (most notably H$\beta$),
the flux is almost independent of shock velocity: such lines are a
good proxy for the mechanical power. Instead, other lines (most
notably [O\,I]\,$\lambda 6300$ and He\,II\,$\lambda 4686$) depend
strongly on $v_{\rm sh}$. Since our models predict the evolution of
$v_{\rm sh}$ as a function of bubble age for a given mechanical power,
we can couple our results to a shock-ionization code, and predict the
evolution of the optical spectrum and hence the characteristic age of
an observed ULXB. This is left to follow-up work.

\begin{table}
\begin{center}
\caption{Shock velocities at characteristic epochs}
\label{tab:shockvel}
\begin{tabular}{llllll}
\hline
\hline
Name & $v_{\rm sh, 0.4Myrs}$ &$v_{\rm sh, 0.5Myrs}$ & $v_{\rm sh, 1Myrs}$ & $v_{\rm sh, 2Myrs}$ \\
\hline
\hline
\texttt{L40v-1.5} & 287 & 191 & 106 & 77  \\
\texttt{L40v-2} & 287 & 191 & 107 & 78 \\
\texttt{L40v-2.5} & 284 & 191 & 110 & 81  \\
\texttt{L40v-3} & 152 & 146 & 120 & 89  \\
\texttt{L39v-2} & 141 & 113 & 69 & 50 \\
\hline
\hline
\multicolumn{5}{l}{$v_{\rm sh, 0.4Myrs}$, $v_{\rm sh, 0.5Myrs}$, $v_{\rm sh, 1Myrs}$ and $v_{\rm sh, 2Myrs}$ are the shock }\\
\multicolumn{5}{l}{velocities after 0.4, 0.5, 1 and 2Myrs of bubble expansion,  }\\
\multicolumn{5}{l}{in $\rm km \, s^{-1}$.}\\

\end{tabular}
\end{center}
\end{table}

\subsubsection{Resolution study}
\label{s.resolution}

A potential weakness of numerical simulations on fixed grids is that
shocks and contact discontinuities (CD) can be under-resolved. We
have already discussed the difficulties in simulating radiative
shocks. Here we discuss the CD. 

Ideally, in the absence of thermal conduction \citep{weaver}, the CD
is described by a step function in both density and temperature, with
no break in the pressure. However, due to numerical diffusion and
finite resolution, simulations are unable to reproduce this sudden
break in the density and temperature. The CD is smoothed out over
a number of cells containing gas at intermediate temperatures and
densities. These intermediate temperatures correspond to the peak of
the cooling function (see Figure~\ref{fig:temp_vs_cooling_2}). Since
the densities in this region are also fairly high, the simulation
would predict a considerable amount of spurious emission from the
region surrounding the CD.

\begin{figure}
\centering
\includegraphics[width=1\linewidth]{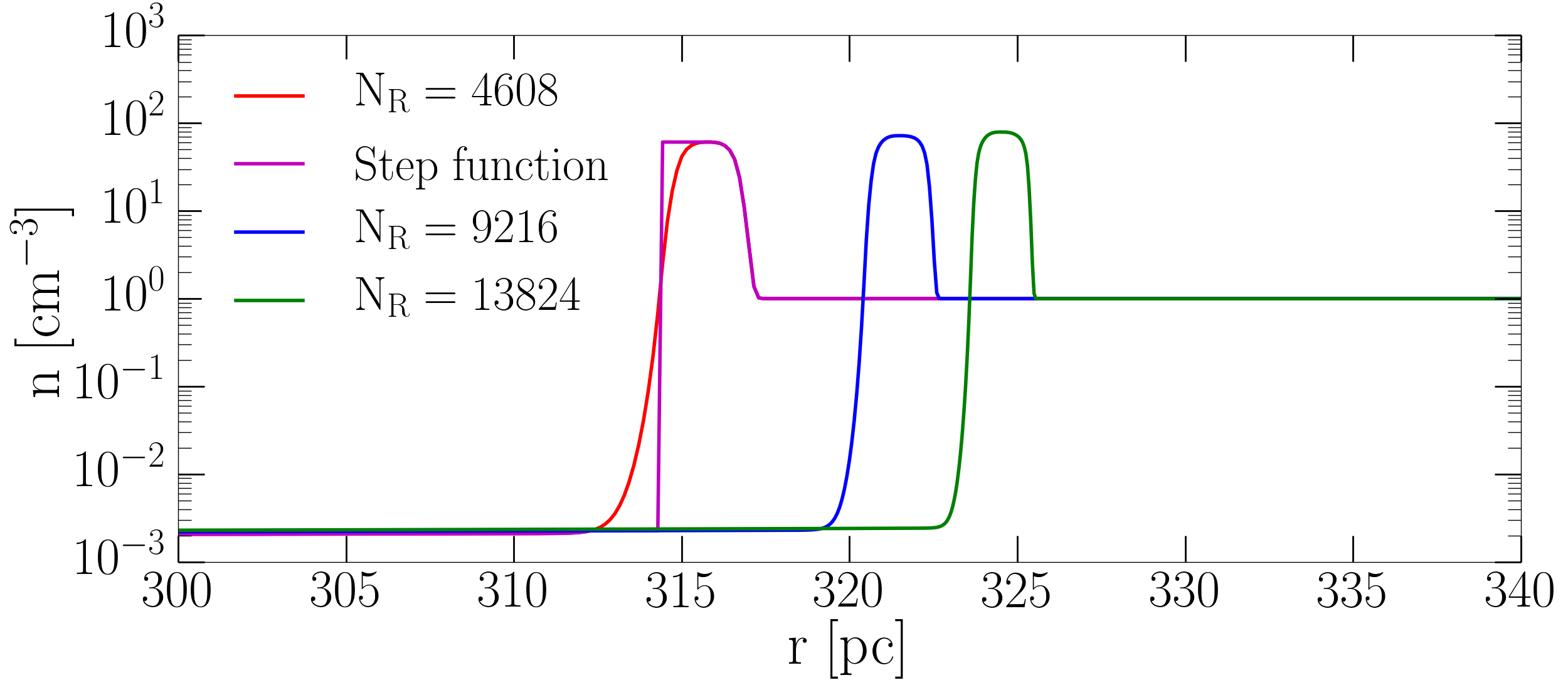}
\includegraphics[width=1\linewidth]{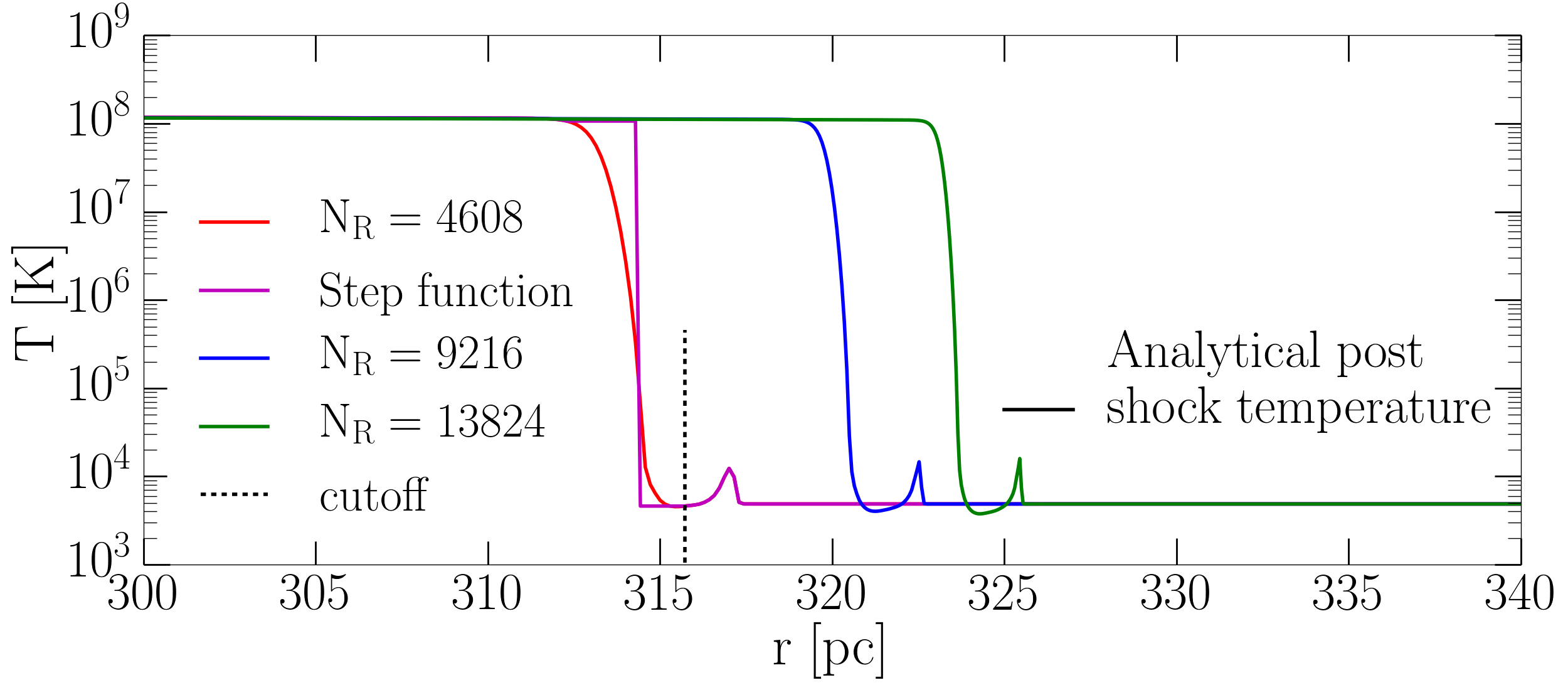}\vspace{-.1cm}
\resizebox{1\linewidth}{.28\textwidth}{\includegraphics[width=1\linewidth]{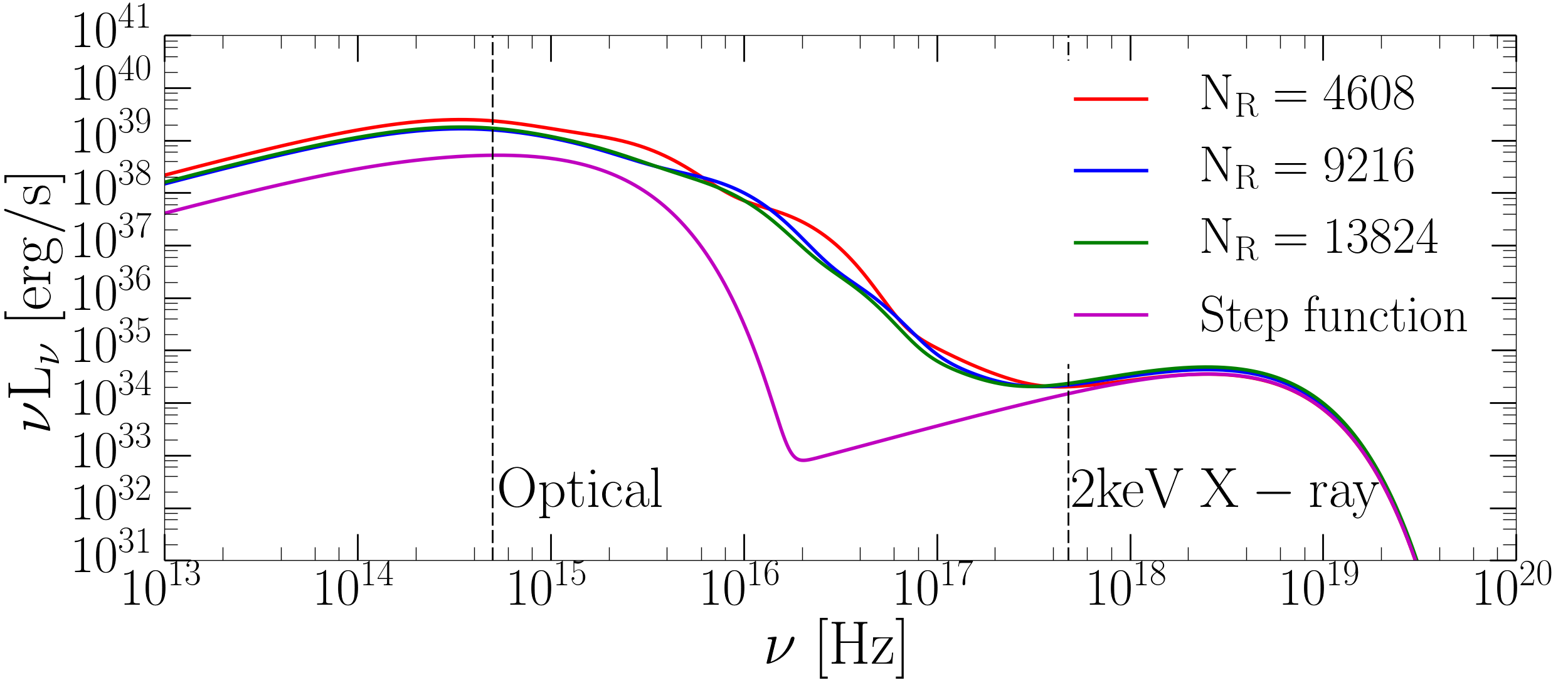}}
\caption{Top and middle panels show the density and temperature
  through the forward shock and the contact discontinuity in the
  fiducial model for three different resolutions: $N_{\rm R} = 4608$ (red
  lines), $N_{\rm R} = 9216$ (blue lines) and $N_{\rm R} = 13824$ (green
  lines). The magenta line shows an idealized model in which the CD is
  taken to be a step function.  The bottom panel displays
  corresponding spectra, using the same colors. The step function CD
  (magenta line) shows considerably less EUV emission.}
\label{fig:all_res}
\end{figure}

Here we compare the profiles and spectra of the forward shock and the
CD for the fiducial model \texttt{L40v-2} during the radiative phase,
obtained with low (4608), medium (9216) and high (13824) spatial
resolution. 
%We compare the post-shock temperature from \texttt{KORAL}
%with our previously calculated analytical value (equation
%\ref{t2}). We investigate how well our simulations are resolving the
%critical regions and how close our resolution study is to convergence.
In Figure \ref{fig:all_res} we show the density and temperature
profiles (top and middle panels) and the computed spectrum (bottom
panel).  Although all profiles are calculated at the same time, they
are offset spatially. This is due to slightly faster expansion in the
higher resolution case during the radiative phase.

Comparison of the density profiles in the top panel of Figure
\ref{fig:all_res} shows a change in CD shape from the lowest
resolution (red, 4608) to moderate resolution (blue line, 9216
cells). The moderate and high resolution (blue and green) models are
more similar, so we conclude that there is not much to be gained by
going above a resolution of $\approx 10000$.

The temperature profile, however, shows that even our highest
resolution case does not resolve the radiative forward shock.  The
temperature of the shocked gas falls well below the analytical
post-shock temperature (marked with the horizontal black line). This
is a well-known problem \citep{hutchings00,creasey11}, and is the
reason why we resorted to an analytical method for calculating the
spectrum of the shocked ISM.

Comparing the spectra (bottom panel) it is apparent that the lowest
resolution case is the brightest in the optical, likely because of
poor resolution near the CD. This is not the case in the X-ray band,
where the three solutions produce emission at the same level, proving
that even the lowest resolution adopted is enough to resolve regions
occupied by the hottest gas.

In Figure~\ref{fig:all_res} we show also a model in which we replace
the numerical CD with a step function (magenta lines). In this case,
the emission in the optical band is lower by a factor of
several. However, the X-ray emission is hardly different from the
other three spectra, confirming that the X-ray emission is not
affected by resolution.  The continuum between the optical and X-ray
band that is seen in the red, blue and green lines is replaced in the
magenta line by a cutoff near $10^{16}$ Hz. Fortunately, this region
of the spectrum is not accessible to observations, so the large
difference is not a problem. As for the optical emission, the magenta
spectrum probably underestimates the true emission because it neglects
the effects of heat conduction at the CD \citep{weaver}, so the true
answer is probably somewhere between this spectrum and the other
three.

\subsection{Parameter study}
\label{s.parameter}

\subsubsection{Expansion}
\label{s.par.expansion}

\begin{figure*}
   \includegraphics[width=1.\columnwidth]{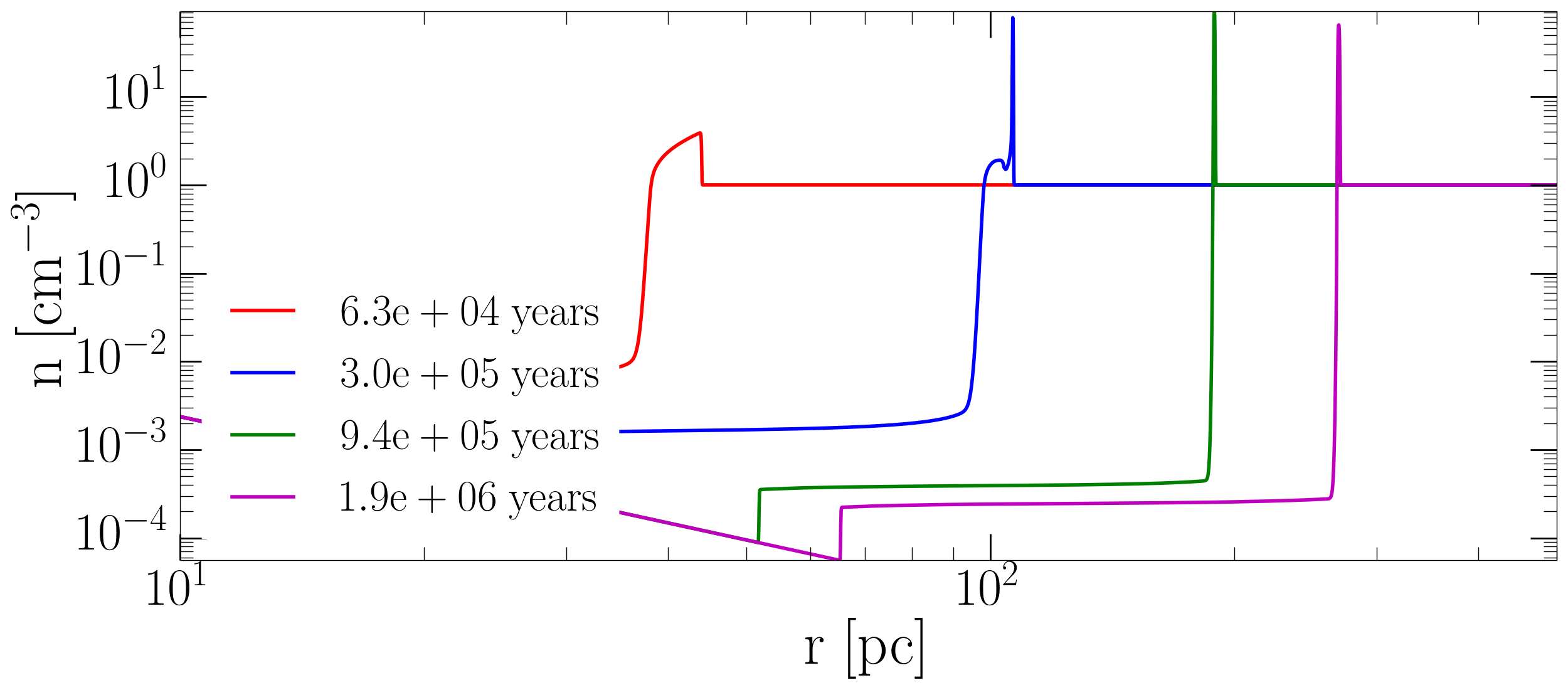}
   \includegraphics[width=1.\columnwidth]{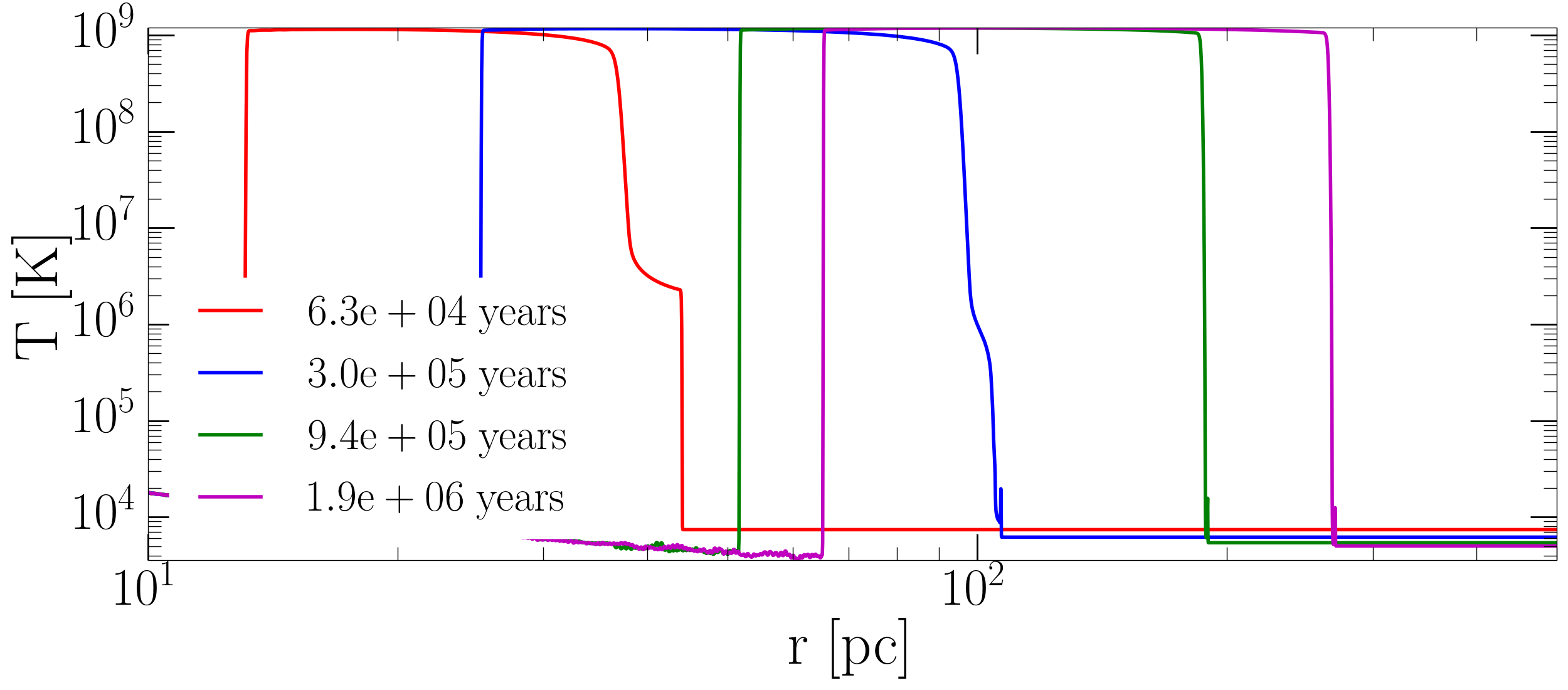}
   \includegraphics[width=1.\columnwidth]{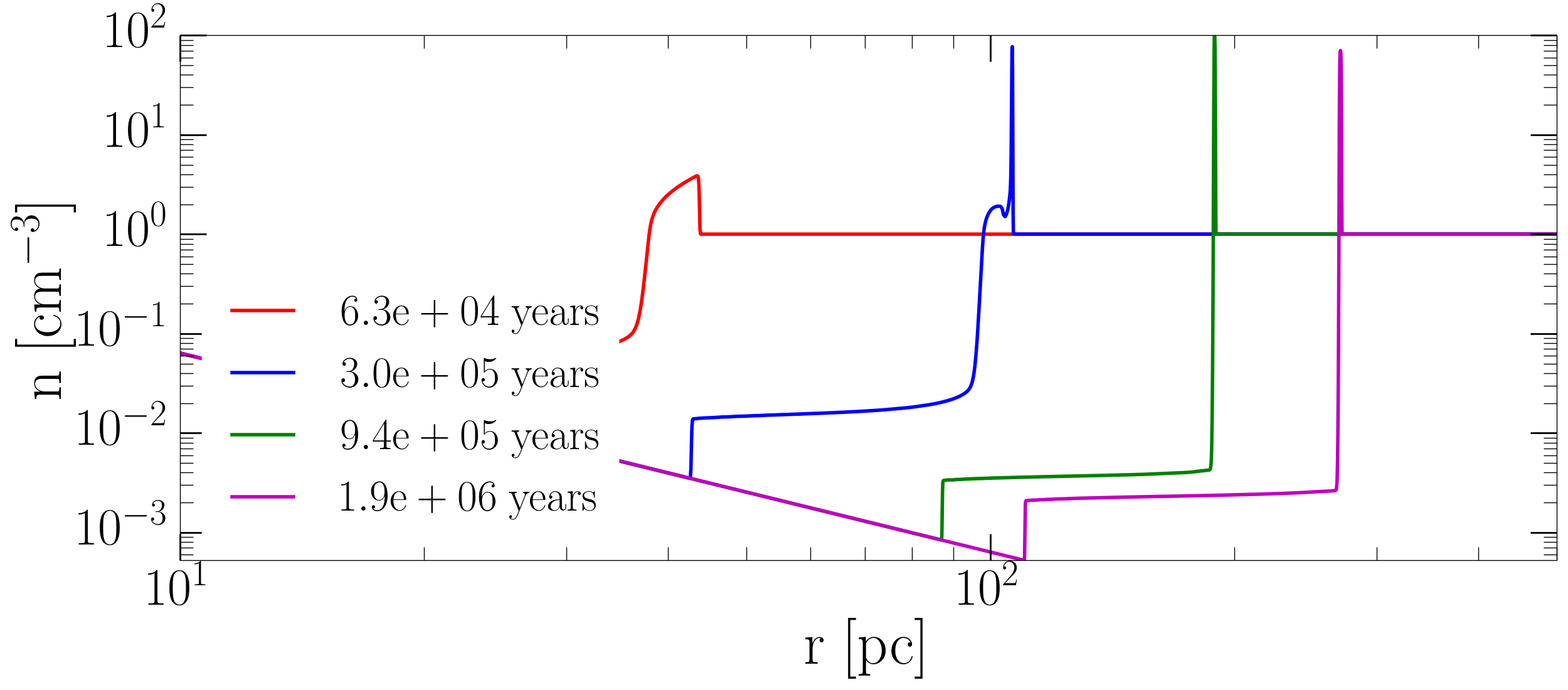}
   \includegraphics[width=1.\columnwidth]{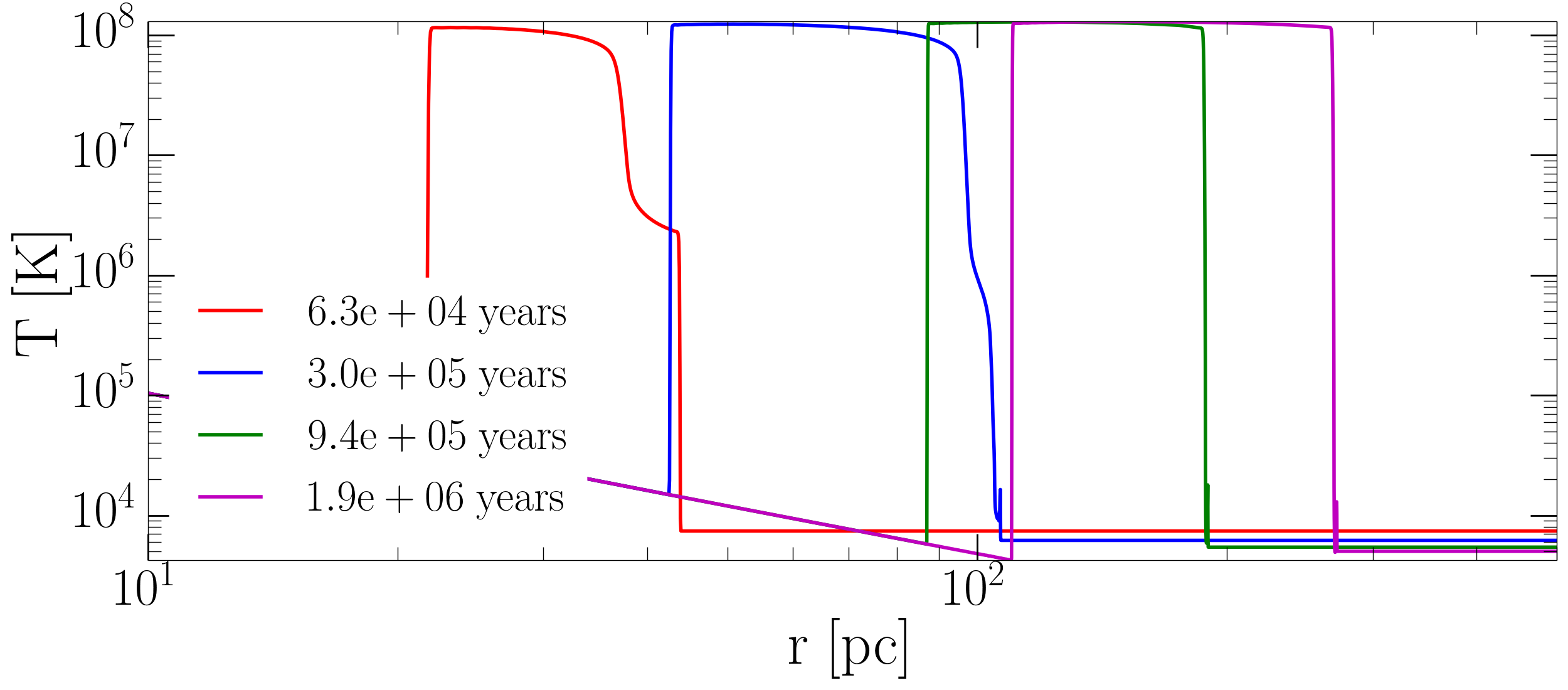}
   \includegraphics[width=1.\columnwidth]{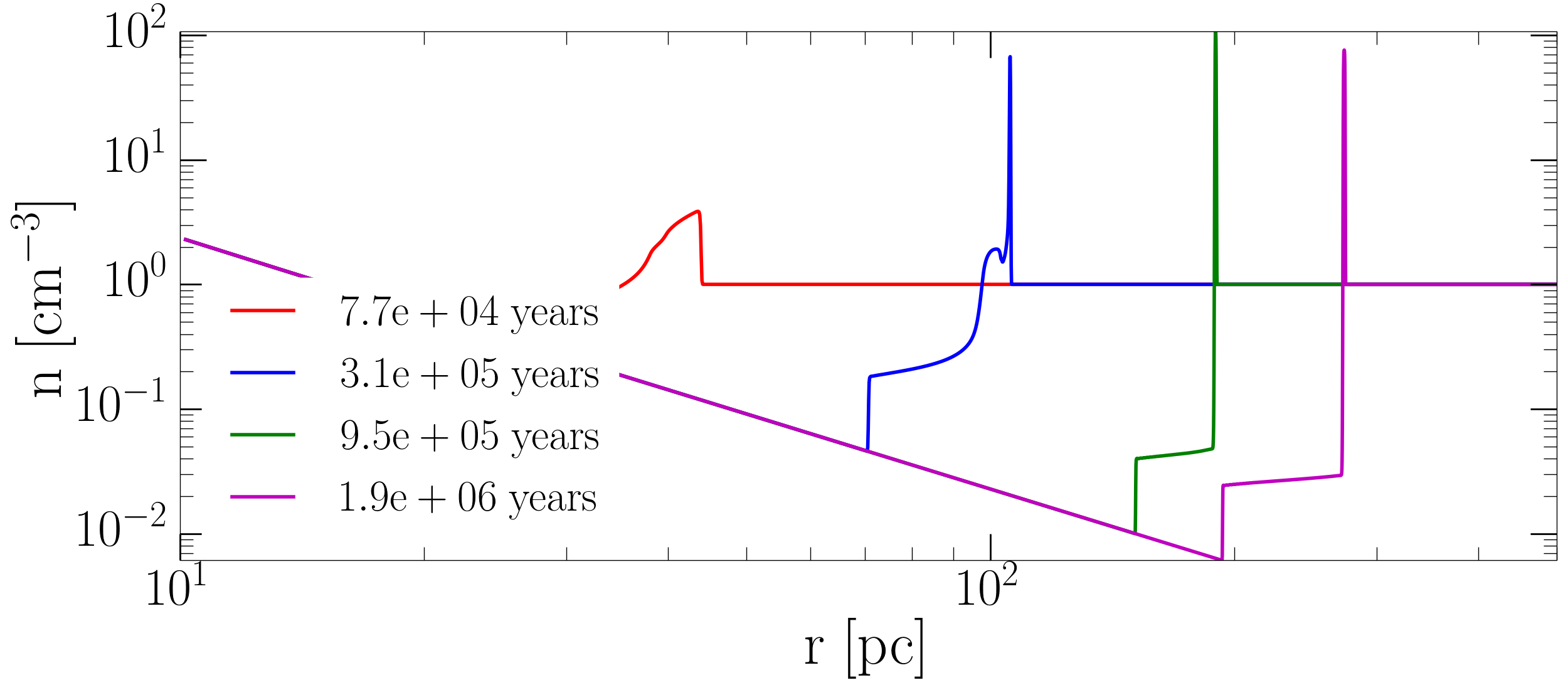}
   \includegraphics[width=1.\columnwidth]{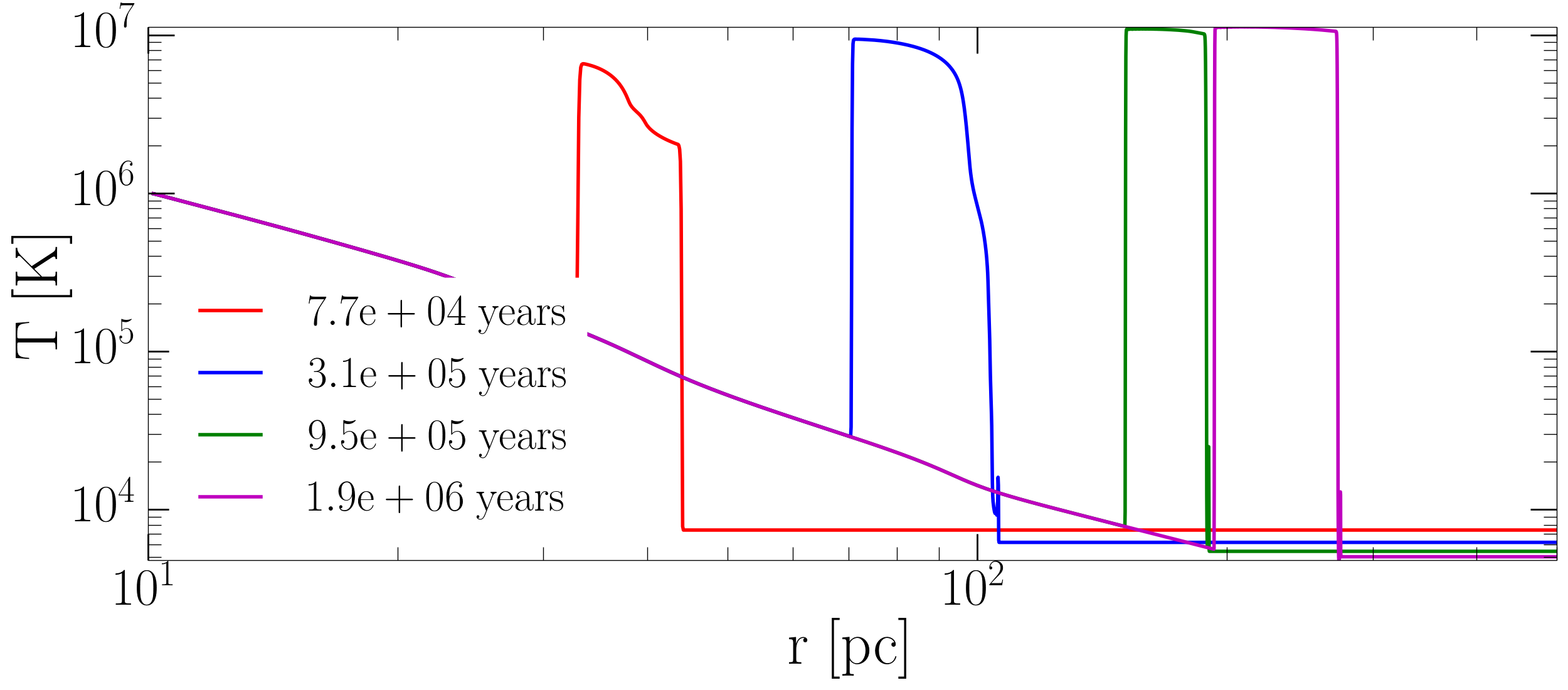}
   \includegraphics[width=1.\columnwidth]{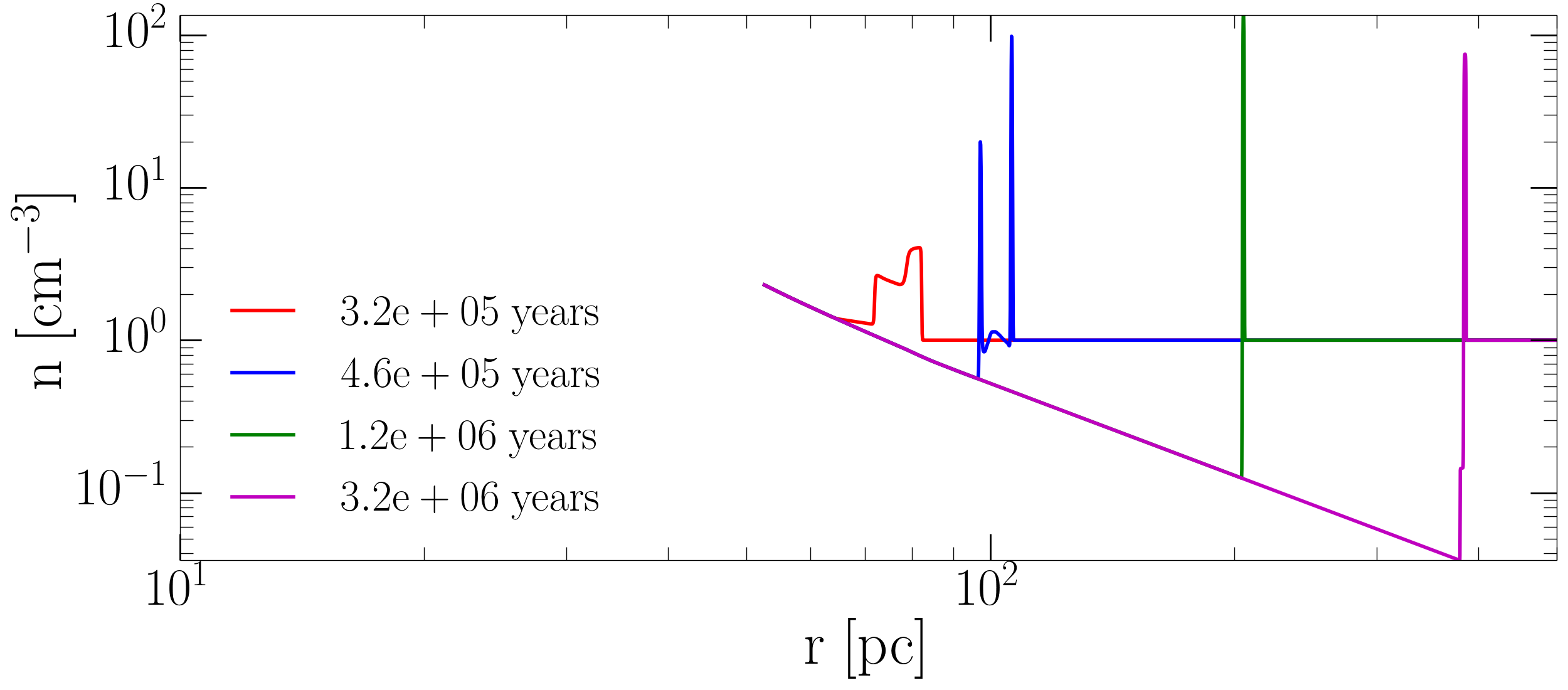}
   \includegraphics[width=1.\columnwidth]{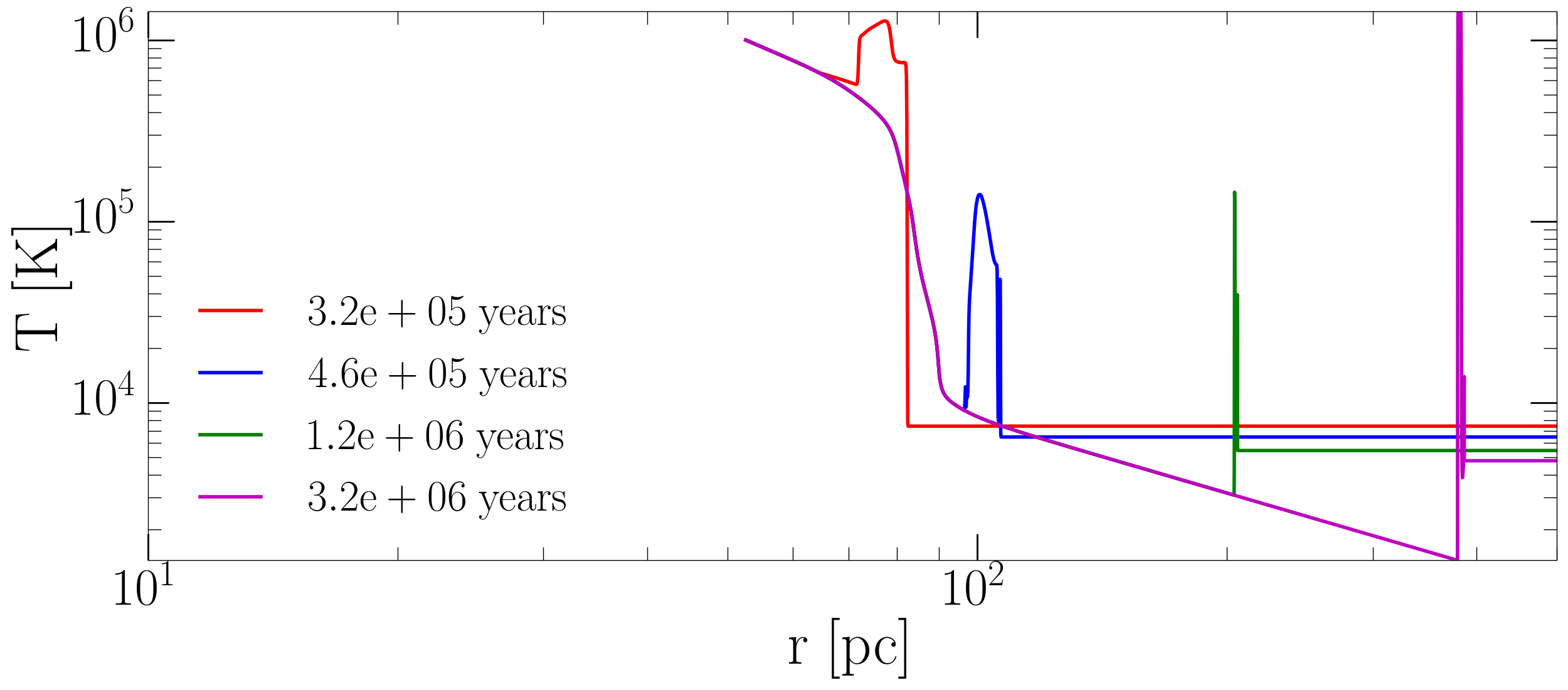}
   \includegraphics[width=1.\columnwidth]{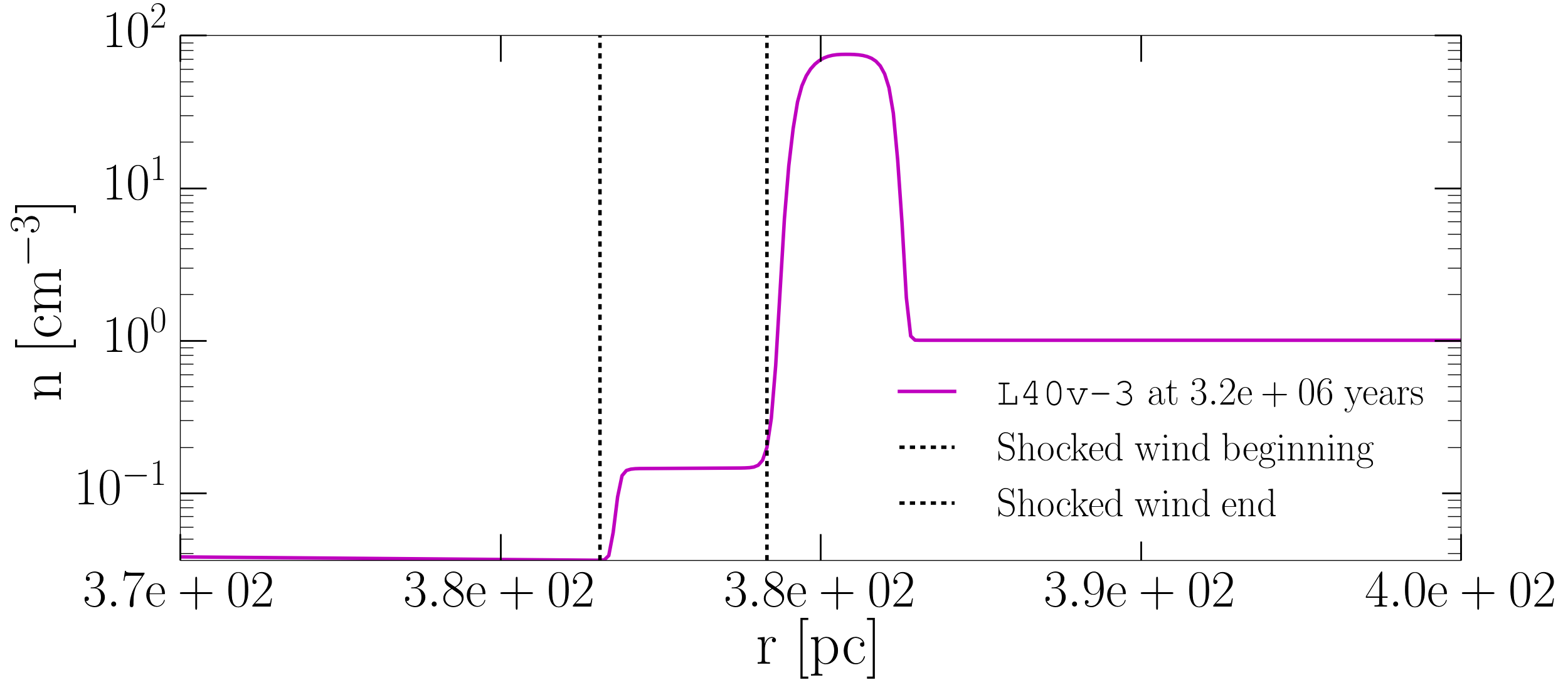}
   \includegraphics[width=1.\columnwidth]{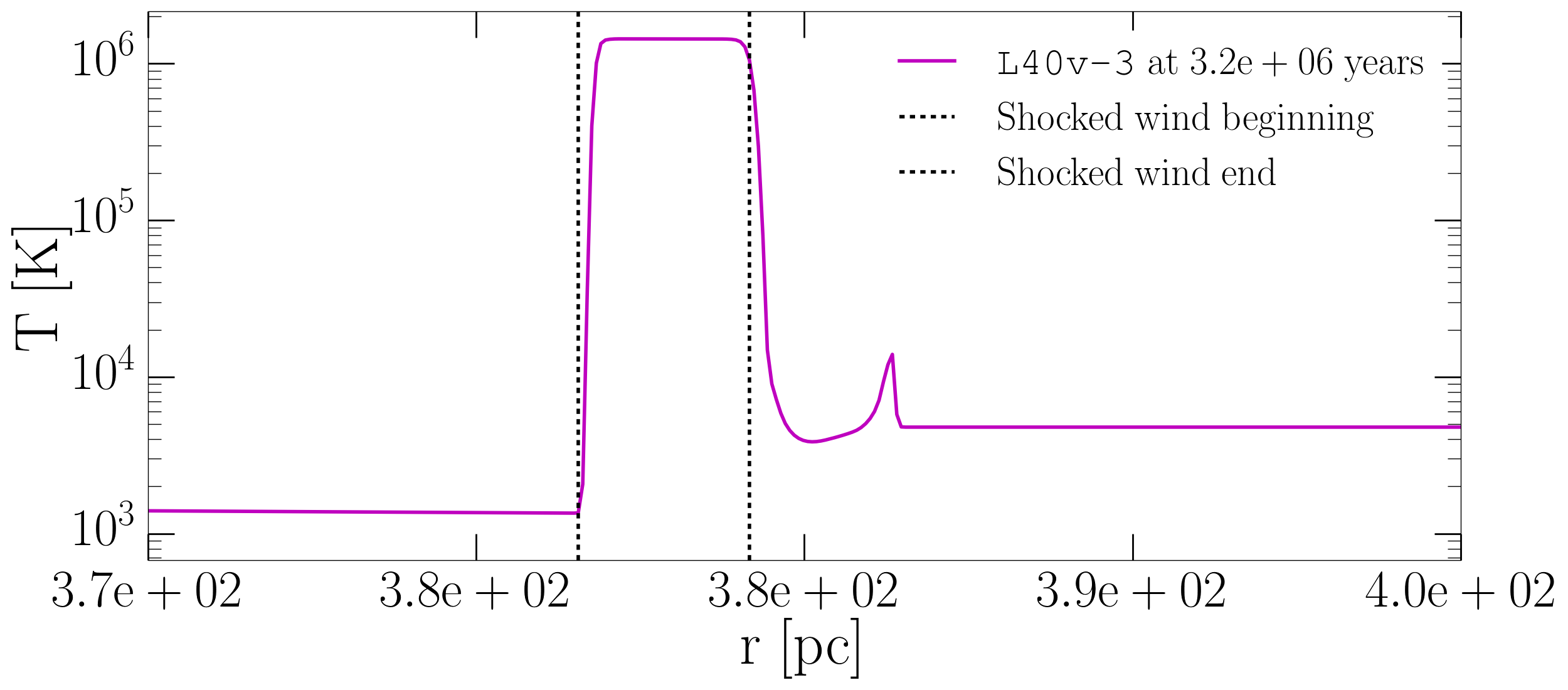}
   \caption{The top four rows show density (left) and temperature
     (right) profiles through ULXBs with wind luminosity $ L_{\rm w} =
     10^{40}  \,{\rm erg\,s^{-1}}$ and wind velocities $ v_{\rm w} = 10^{-1.5}  \,
     c$, $ v_{\rm w} = 10^{-2} \, c$, $ v_{\rm w} = 10^{-2.5}  \,c$ and $
     v_{\rm w} = 10^{-3}  \,c$, respectively (from top to bottom). Rows
     1-3 show the standard three phases: adiabatic expansion (red),
     transition from adiabatic to radiative expansion (blue), and
     radiative expansion (green and magenta). All shocks seem to
     propagate at comparable speeds. However, as the wind velocity
     decreases, so does the extent of the shocked wind envelope (rows
     1-3). In row 4 a small envelope of shocked wind and swept up ISM
     is visible only at early times (red line). Subsequently, the
     shocked wind and ISM both simultaneously collapse into thin
     shells due to efficient cooling (blue line, see discussion in
     Section~\ref{s.parameter}). The snapshots in the fourth row
     panels are taken at slightly different times compared to the
     other three models, to better illustrate the characteristic
     properties of bubble expansion in this regime. In the bottom row,
     the late stage of the model with $ v_{\rm w} = 10^{-3}  \,c$ is
     shown, and illustrates the appearance of an adiabatic shocked
     wind at late times.}
\label{dens_and_temp}
  \end{figure*}

In Figure \ref{dens_and_temp} we show density and temperature profiles
corresponding to the adiabatic, transition and radiative phases for
models with $ L_{\rm w} = 10^{40} \, {\rm erg\,s^{-1}}$ and wind velocities $
v_{\rm w} = 10^{-1.5} c$, $ v_{\rm w} = 10^{-2} c$, $ v_{\rm w} =
10^{-2.5} c$ and $ v_{\rm w} = 10^{-3} c$ (models \texttt{L40v-1.5},
\texttt{L40v-2}, \texttt{L40v-2.5}, \texttt{L40v-3}).

All four of these ULXB models begin the transition from the adiabatic
to the radiative phase at radius $ R \approx 100$\,pc. There is a clear
trend in panels 1-3: As the wind velocity decreases, the density of
the shocked wind envelope increases (because the density of the
unshocked wind itself increases) while the temperature of the shocked
wind decreases (because of the lower wind velocity). The shocked wind
temperatures are a few $10^9 \,$K, $10^8 \,$K and $10^7 \,$K for models
\texttt{L40v-1.5}, \texttt{L40v-2}, \texttt{L40v-2.5},
respectively. The properties of the shocked ISM are identical in all
three cases throughout the ULXB evolution, as are the shock expansion
rates.

Panel 4 ($ L_{\rm w} = 10^{40}$, $ v_{\rm w} = 10^{-3} c$) shows a
qualitative break from the previous pattern. The shocked wind and ISM
become radiatively efficient at roughly the same time (again at radius
$ R \approx 100$ \,pc), and both zones collapse to a thin, dense shell of
cold gas with $\rm T \approx 10^4 \, K$. This model behaves differently
because, given the low wind velocity $ v_{\rm w} = 0.001c$, the gas
crossing the reverse shock heats up to temperatures below $\rm T_{\rm
  w} \approx 10^6 \, K$, where the cooling function peaks (Figure
\ref{fig:temp_vs_cooling_2}), causing the shocked wind to collapse
through catastrophic cooling.  At later stages (purple lines in the
bottommost panels of Figure \ref{dens_and_temp}) a radiatively
inefficient, hot wind layer begins to develop again behind the contact
discontinuity. This happens despite the low shocked wind temperature
because of decreasing gas density in this regime --- the density is no
longer large enough to provide efficient cooling.

\subsubsection{Radiative properties}
\label{s.par.radiative}

\begin{figure}
\centering
\resizebox{1\linewidth}{.27\textwidth}{\includegraphics[width=1\linewidth]{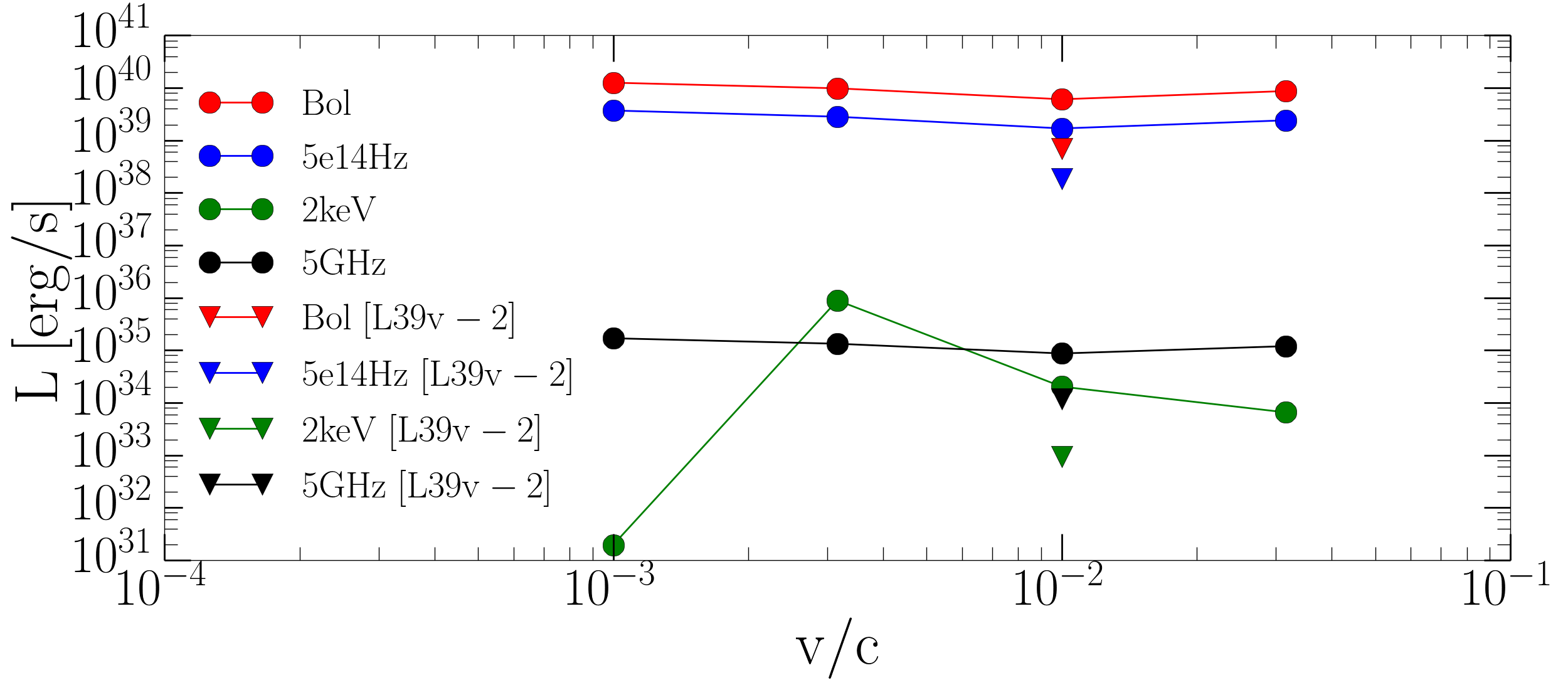}}
\caption{Bolometric, optical ($5\times10^{14}$ Hz), radio (5GHz) and
  X-ray (2\,keV) emission of the ULXB as a function of wind
  velocity. All luminosities are taken from the radiative phases of
  the ULX bubble at $ \approx 2\,$Myrs. For $v_{\rm w}=10^{-3}c$ the
  shocked wind cools efficiently and no significant X-ray emission is
  observed.}
\label{fig:L_vs_wind_velocity_with_Le39}
\end{figure}

Figure \ref{fig:L_vs_wind_velocity_with_Le39} shows the velocity
dependence of the total luminosities of the ULXB models during the
radiative phase.  The bolometric luminosity stays approximately
independent of the wind velocity, except for the model with $ v_{\rm
  w} = 10^{-3}c$, in which the bolometric luminosity exceeds $10^{40}  \, {\rm erg\,s^{-1}}$ slightly (see below). The radiation is emitted
predominantly in the UV/optical band, where the luminosity is of order
$10^{39}  \, {\rm erg\,s^{-1}}$ for all the models, independent of the wind
velocity. Similar to the optical, the free-free radio emission does
not show significant dependence on wind velocity.

The X-ray emission comes primarily from the shocked wind. It is at
most $10^{36} \, {\rm erg\,s^{-1}}$ (in model \texttt{L40v-2.5}), and can be as
low as $10^{31} \,{\rm erg\,s^{-1}}$ (in model \texttt{L40v-3}). Although at the
highest wind velocities, e.g., $ v_{\rm w} = 0.03c$ and $v_{\rm w} =
0.01c$, the volume of the shocked wind is largest and the temperature
is highest (see Figure~\ref{dens_and_temp}), the X-ray luminosities of
these models are low.  The reason is that the amount of X-ray emission
is a trade off between shocked wind volume, temperature and density.

The emissivity at X-ray temperatures scales as $\epsilon \sim n^2
T^{1/2}$, hence the X-ray luminosity $L_X$ of the bubble scales as
\begin{equation}
 L_X = \epsilon V \sim n^2T^{1/2}V,
\end{equation}
where $V$ is the volume of the shock-heated gas.  Going from model
\texttt{L40v-1.5} ($v_{\rm w} = 0.03c$) to model \texttt{L40v-2.5}
($v_{\rm w} = 0.003c$), the temperature decreases by two order of
magnitude, but the density increases by two orders of magnitude. Since
density has a much stronger effect on the emissivity, the result is
that the model with the slower wind has a substantially larger $L_X$.
This trend continues until the wind velocity is so low that the gas
temperature falls below the X-ray band. This is the case with model
\texttt{L40v-3}, which has almost no X-ray emission.  The net result
is that $L_X$ is largest for $ v_{\rm w} = 0.003c$, which is still hot
enough for X-ray emission ($>10^6 \,$K), and has the largest density
among the models that satisfy this requirement.

In model \texttt{L40v-3}, the shocked wind
is radiatively efficient and the gas cools considerably.  Therefore,
the model radiates predominantly in the UV and optical. This emission
is responsible for the much higher bolometric luminosity of this
model.  In fact, the radiative luminosity is slightly {\it above}
$10^{40} \, {\rm erg\,s^{-1}}$ (the injected wind kinetic luminosity) during the
radiative phase (Figure \ref{fig:L_vs_wind_velocity_with_Le39}).  The
extra luminosity results from the emission of thermal energy
accumulated during the adiabatic phase.

\section{Discussion}
\label{s.discussion}

\subsection{Range of validity of the simulations}
\label{caveats}

As we mentioned in Section 1, ULXs affect the surrounding ISM both via
X-ray photoionization and via shock-ionization. In this work, we
modelled the case of shock-ionized nebulae formed by continuous
outflow.  Our simulations are based on a number of simplifying
assumptions. To start with, we simulated the expanding nebulae
assuming perfect isotropy. This is never satisfied in reality. Although the outflow from the accreting system will carry an imprint
of the accretion flow geometry, the ISM is never a
uniform medium, and fluctuations in its density will affect the
expansion of the nebula, making it highly non-isotropic. 

Radiative shocks are known to be subject to thermal and thin-shell instabilities \citep{Vishniac1983,Chevalier1982}, and we therefore tested our model for Rayleigh-Taylor (R-T) instabilities by running 2-D simulations. We did not detect any instabilities, though this may be due to the low spatial resolution we employed. Previous studies such as \cite{Porth2014} did observe the development of R-T filaments in adaptive mesh refinement MHD simulations of pulsar wind nebulae, but those simulations used much higher (> 1 order of magnitude) resolution compared to ours. If R-T instabilities occur, we might expect the cold ISM to mix with the hot shocked wind in the ULXB. The mixing could cool the shocked wind enough to put it in the temperature regime of effective cooling ($\lessapprox 10^6$\,K), thus shortening the adiabatic expansion phase of the ULXB. In future work high resolution 2-D simulations should be considered to test the effects of instabilities on emission spectra and luminosities of ULXBs in the radiative phase. \color{black}

In the present work we also neglected the presence of a relativistic,
magnetized and collimated BH jet, predicted by MHD simulations as one
of the hallmarks of the super-Eddington accretion regime
\citep{sadowski+koral2,rs2,rs1,rs3,narayan17}. Relativistic jets have
been detected or inferred so far for a small number of (candidate)
super-Eddington stellar-mass BHs: most notably SS\,433
\citep{rs4,rs5}, Holmberg II X-1 \citep{cseh}, and the ultraluminous
supersoft source in M\,81 \citep{rs6}. It is still unknown whether all
super-critical BHs have jets, and whether the jet power depends on BH
spin and on the magnetic configuration near the BH horizon. A jet
carries most of its energy in a small opening angle, leading to the
formation of an elongated nebula with lobes and hot spots (as observed
in the ULX bubble NGC\,7793 S26, \cite{Pakull2010} and
\cite{Soria2010}). Even when jets are present, MHD simulations of BH
accretion predict that a comparable amount of power is carried by
wide-angle, slower outflows \citep{parameter}: in that case, the
likely effect on the ISM is a more spherical bubble inflated primarily
by the slower wind, with jet-inflated lobes sticking out of the bubble
(similar to the morphology of the SS\,433/W50 nebula). Our
spherically-symmetric model would then more properly apply only to the
spherical component of the bubble. 
%however, our results provide useful
%insights for a qualitative understanding also of the jet-inflated
%lobes.

Since we are primarily concerned with wide-angle, massive, slower
outflows, we limited our simulations to wind velocities between
$0.001c$ and $0.03c$; the narrow jet component is the only part of the
outflow that can reach speeds $\gtrsim 0.1c$. Physically, the lower
limit of our wind velocity interval corresponds to a threshold where
the shocked wind region becomes radiatively efficient and collapses to
a thin, dense shell with $T \sim 10^4$ K. The upper limit of our
velocity interval does not correspond to any physical thresholds; it
is the escape velocity of outflows launched from a disk radius
$\approx 2000 R_{\rm g}$.

Another consequence of our focus on wind-inflated rather than
jet-inflated bubbles is that the radio luminosity calculated in
Section 3.1.2 includes only the contribution from thermal free-free emission. If a jet is
present, we expect also synchrotron emission from relativistic electrons in the
lobes and hot spots. For powerful jets, such a component may dominate
over the free-free component (as is the case for example in
NGC\,7793-S26: \cite{Soria2010}). 
%Thus, our models provide a useful
%baseline level of radio luminosity from the shocked gas: if
%observations of ULXBs show a much higher radio luminosity, it could be
%taken as evidence that the bubble contains an additional population of
%relativistic electrons from a jet. The spectral index of the observed
%radio emission will also help us determine whether it is dominated by
%synchrotron or bremsstrahlung.
Since we ignore synchrotron radiation, our radio luminosity estimates should be viewed as lower limits to the observable radio emission. If the radio comes primarily from a jet ($v \ge 0.1\,c$) then, extrapolating \cite{Cavagnolo2010} to the typical jet power in ULXs,  we estimate that
\begin{equation}
L_{\rm radio} \sim 10^{-5} \times P_{\rm jet}.
\end{equation} 
We therefore expect synchrotron radio luminosities of order $\sim 10^{35} \rm \, erg\, s^{-1}$ in jet-dominated ULXBs. This is supported by radio observations of the ULX bubbles S26 in NGC7793 \citep{Soria2010} and MQ1 in M83 \citep{Soria2014}. 
On the other hand, ULXBs such as Ho IX X-1 and NGC1313 X-2 are just as large and optically luminous, but have low radio luminosities, consistent with just thermal bremsstrahlung emission. 
It is likely that radio-loud ULXBs are inflated by jets, and radio-quiet ULXBs by winds: for the same amount of kinetic power, jet shocks may be more efficient than winds in accelerating non-thermal electrons. If so, then our calculations give a baseline radio luminosity in the absence of jets, and any synchrotron jet component would be on top of that.
 \color{black}

We did not include conduction in our simulations. \cite{weaver}
calculated the evaporation of shocked ISM into the shocked wind layer
due to conduction, and concluded that 60\% of the mass in the shocked
wind comes from evaporation across the CD. Conduction can therefore
alter the shape of the CD, and the observed spectral properties. The
expansion speed of the shock would also be affected. We implicitly
assumed that conduction is negligible.

We also do not consider possible non-thermal support by cosmic rays accelerated at the shock or by magnetic fields. \cite{Castro2011} compare the evolution of supernova remnants with and without efficient cosmic ray production and show that cosmic ray production results in higher compression ratios and lower post-shock temperatures, and shocks with smaller radius and speed, as well as lower intensity X-ray thermal emission. These observational properties can be fit to a model with a different set of parameters, in particular a much lower explosion energy, compared to models that include cosmic ray production. In future work we could account for the effects of cosmic ray production on the emitted X-ray spectrum in ULXBs by using a non-equilibrium ionization collisional plasma model, contained in \texttt{XSPEC} \citep{9781583817643}.

We further treated the gas behind the front shock as a single temperature plasma, rather than tracking proton and electron temperatures separately. Recent work on AGN outflows \citep{faucher} has shown that proton cooling timescales can exceed electron cooling timescales significantly, and thus increase the parameter range over which such outflows are adiabatic. It would be interesting to rerun the simulations reported here using separate cooling functions for electrons and protons. \color{black}

\subsection{Diagnostics of wind speed}
\label{s.implications}

The standard model of the expansion of a wind-inflated nebula predicts
that the properties of the shocked ISM depend only on the mechanical
luminosity of the outflow. In particular, shell temperature and
luminosity do not depend on the density or velocity of the
outflow. The evolution of the swept-up ISM shell in our simulations is
consistent with this picture.

The properties of the gas that crosses the reverse shock and forms the
hot and tenuous expanding envelope behind the CD are, on the other
hand, sensitive to the BH wind properties. For higher wind speeds (and
therefore, at a fixed mechanical luminosity, for lower wind
densities), post-shock temperatures are larger. The slowest wind speed
in our simulations corresponds to the lowest temperature, but also to
the highest integrated emission
(Fig.~\ref{fig:L_vs_wind_velocity_with_Le39}) because of the shape of
the cooling function between $T \sim 10^5$--$10^8$ K.

By simulating the bubble expansion numerically using a realistic cooling
function we were able to extend this classical picture to the regime
where the post-shock wind temperature becomes low,
bound-free emission becomes effective and the envelope of the shocked
wind cools efficiently and collapses to a thin layer, just behind the
swept-up ISM shell. In this case the radiative properties of the
shocked wind change dramatically, leading to efficient emission in
optical wavelengths rather than in X-rays.

This peculiar dependence of the X-ray emission and of the X-ray/radio
luminosity ratio on the outflow properties
(Fig.~\ref{fig:L_vs_wind_velocity_with_Le39}) could enable us, in
principle, to disentangle outflow rate and velocity. For a given
nebula it is relatively straightforward to estimate the mechanical
power required to inflate the observed cavity. If, in addition, limits
can be placed on the X-ray emission from within the bubble, then the
unshocked wind velocity may be recovered. This would in turn constrain
the outflow densities. Both parameters provide key constraints to MHD
simulations of the accretion flow in the innermost region around the
BH. The dependence of the X-ray spectrum on wind speed is even
stronger if we use more realistic thermal-plasma emission models ({\it
  e.g.}, the APEC model: \cite{rs7}), including metal lines in
addition to free-free emission. In practice, diffuse
X-ray luminosity at the level of $L_{\rm X} \lesssim 10^{36} \, {\rm erg\,s^{-1}}$ from ULXBs located at least several Mpc away would require exposure times above 2\,Ms to be detected.

\section{Summary}
\label{s.summary}

We performed one-dimensional simulations of expanding shock-ionized
bubbles powered by ULX outflows. We accounted for realistic radiative
losses by implementing a cooling function that includes both free-free and bound-free opacities. We studied the properties of the
nebular expansion as a function of the mechanical luminosity and
velocity of the ULX outflow. We found that:

(i) Consistent with standard analytical models of expanding,
wind-driven bubbles, the swept-up, shocked ISM quickly enters the
radiative phase and collapses into a thin shell, with peak radiative
emission in the optical band and integrated luminosity comparable to
the mechanical output of the ULX. The properties of the shocked ISM
depend only on the input power (via the shock velocity $v_{\rm sh}$),
and not on the velocity or density of the wind.

(ii) For models \texttt{L40v-1.5}, \texttt{L40v-2} and
\texttt{L40v-2.5}, the shock velocity decreases with time, from
characteristic values $v_{\rm sh} \approx 300\,$km$\,$s$^{-1}$ at an age
$\approx 10^5\,$yr to $v_{\rm sh} \approx 100  \,$km$ \,$s$^{-1}$ at an age
$\approx 10^6\,$yr (Table \ref{tab:shockvel}).  The value of
$v_{\rm sh}$ is the main parameter required to predict realistic
optical/UV line emission spectra of the bubble as a function of
time. The range of $v_{\rm sh}$ found in our models is consistent with
those inferred from the line spectra observed in large ULXBs
\citep{pakull02,pakull,moon}. By modelling the evolution of $v_{\rm
  sh}$ in time, we provide an independent method to determine the age
of a ULXB, via its optical line ratios.

(iii) An envelope of shocked wind develops behind the contact
discontinuity, again in agreement with the standard models. Contrary
to the shocked ISM, the properties of the shocked wind do depend on
the input parameters (wind speed and density); higher wind velocities
correspond to lower densities and higher temperatures. The shocked
wind does not cool efficiently if the wind velocity $v_{\rm w}\gtrsim
0.003c$, leading to temperatures $\sim 10^7$--$10^8\,$K and X-ray
emission. For an input mechanical power $L_{\rm w} = 10^{40}\, {\rm erg\,s^{-1}}$, the characteristic bremsstrahlung luminosity at 2 keV is
negligible for wind speeds $v_{\rm w} \approx 0.001c$, peaks at
$L_{\rm X} \approx 10^{36}\,{\rm erg\,s^{-1}}$ for $v_{\rm w} \approx
0.003c$ ($T \approx 10^7$ K), and decreases to $L_{\rm X} \approx
10^{34}\,{\rm erg\,s^{-1}}$ for $v_{\rm w} \approx 0.03c$ ($T \approx 10^9\,$K).

(iv) For a given input mechanical power, the free-free radio
luminosity (integrated over the whole volume of the bubble) is almost
independent of wind velocity; instead, it is a function of bubble
age. For a mechanical power of $10^{40}\,{\rm erg\,s^{-1}}$, the 5-GHz
luminosity is $\lesssim 10^{34}\,{\rm erg\,s^{-1}}$ in bubbles younger than
$\approx 3 \times 10^5$ yr, and jumps to $\approx 10^{35}\,{\rm erg\,s^{-1}}$ after that. At a characteristic distance of 5 Mpc, this
corresponds to a 5-GHz flux density $\lesssim$70 $\mu$Jy for a younger
bubble, and $\approx$700 $\mu$Jy for an older bubble. What we
calculated here are the minimum levels of radio luminosity expected
for a ULXB: they include only free-free emission, not the
optically-thin synchrotron emission from the possible jet lobes and hot
spots. A small sample of observed ULX jets with powers $\approx
10^{40}\,{\rm erg\,s^{-1}}$ suggests that the jet-powered synchrotron
emission at 5 GHz is also $\sim 10^{35}\,{\rm erg\,s^{-1}}$
\citep{Soria2010, Soria2014, cseh12}.

(v) If the disk outflow velocity $v_{\rm w} \lesssim 0.001c$, the wind
that crosses the reverse shock heats up only to moderate temperatures,
low enough to allow for efficient bound-free cooling. As a result, the
shocked wind cools down and collapses, similarly to the behaviour of
the shocked ISM. Instead of having a thin shell of ISM surrounding an
extended envelope of hot shocked wind, now there are two thin and cool
shells, both emitting efficiently in the optical band. During this
phase, the total luminosity is approximately equal to the mechanical
energy injection rate; there is no X-ray emission owing to the
efficient cooling and low temperatures of both the shocked ISM and
shocked wind layers.

\section{Acknowledgements}

MS and RN were supported in part by NSF grant AST1312651. RN was also
funded by the Black Hole Initiative at Harvard University, which is
supported by a grant from the John Templeton Foundation. AS
acknowledges support from NASA through Einstein Postdoctoral
Fellowship number PF4-150126 awarded by the Chandra X-ray Center,
which is operated by the Smithsonian Astrophysical Observatory for
NASA under contract NAS8-03060.  TPR acknowledges support from STFC as
part of the consolidated grant ST/L00075X/1. The authors acknowledge
computational support from NSF via XSEDE resources (grant
TG-AST080026N).

\bibliographystyle{yahapj}
\bibliography{myrefs}

\end{document}